\documentclass{aa}
\usepackage{natbib}
\usepackage{graphicx}

\bibpunct{(}{)}{;}{a}{}{,}

\begin{document}

\title{Impact of Supernova feedback on the Tully-Fisher relation}

\author{M. E. De Rossi \inst{1,2}
        \and P. B. Tissera \inst{1,2}
        \and S. E. Pedrosa\inst{1,2}
        }

\offprints{M. E. De Rossi}

\institute{Consejo Nacional de Investigaciones Cient\'{\i}ficas y T\'ecnicas, CONICET, Argentina\\
\email{derossi@iafe.uba.ar}
\and Instituto de Astronom\'{\i}a y F\'{\i}sica del Espacio, Casilla de Correos 67, Suc. 28, 1428, Buenos Aires, Argentina\\
\email{patricia@iafe.uba.ar,supe@iafe.uba.ar}
}

\date{Received / Accepted}

\abstract {Recent observational results  found a bend in the Tully-Fisher Relation in such a way that low mass systems lay below the linear relation described by more massive galaxies.} 
{We intend to investigate the origin of the observed features in the stellar and baryonic Tully-Fisher relations and analyse the role played by galactic outflows on their determination.}
{Cosmological hydrodynamical simulations which include Supernova feedback were performed in order to
  follow the dynamical evolution of galaxies.} 
{We found that Supernova feedback is a fundamental process  in order  to reproduce
the observed trends in the stellar Tully-Fisher relation. Simulated slow rotating systems
tend to have lower stellar masses than those predicted by the linear fit to the massive end 
of the relation,
consistently with observations.  
This feature is not present  if Supernova feedback is turned off. 
In the case of the baryonic Tully-Fisher relation, we also detect a weaker tendency
for smaller systems to lie below the linear relation described by larger ones.
This behaviour arises  as a result of the more efficient action
of Supernovae in the regulation of the star formation process and in the triggering of powerful galactic
outflows in shallower potential wells which may heat up and/or expel  part of the
gas reservoir.
} {}


\titlerunning{SN feedback and the Tully-Fisher relation}
\authorrunning{De Rossi et al.}

\maketitle

\section{Introduction}

The origin of the Tully-Fisher relation \citep{tf77} has been analysed by numerous 
observational and theoretical works since it provides important constraints for 
galaxy formation models \citep[e.g.][]{vd98, mo98, avila09}.
Originally defined as the relation between the luminosity and the rotation velocity 
for spiral galaxies, it is now accepted that this is a proxy of a more fundamental 
one between stellar mass
and rotation velocity \citep[e.g.][]{cons05, flo06, cres09}
 or even between baryonic mass and rotation velocity 
\citep[e.g.][]{bdj01, ver01, guro04, mcga05, puech10}  of the form $M \propto V^{\alpha}$. 
Observational evidence suggests the existence of a break in the stellar Tully-Fisher relation (sTFR) at 
a rotation velocity of $ \sim 90 \, {\rm km \, s^{-1}}$ \citep{mcg00}.  
Even more, recently \citet{mcg10} also reported a break  in the baryonic Tully-Fisher relation (bTFR) but at a lower 
velocity.

Assuming that galaxies formed in virialized haloes and that the fraction of gas transformed into stars 
is independent of  halo mass,  a relation between stellar mass and circular velocity  with a slope of
$\alpha \sim 3$ is directly obtained \citep{wf91, mo98}. 
This theoretical prediction is reproduced in cosmological simulations when non effective mechanisms
 to regulate the gas cooling and star
formation activity  are introduced \citep{tis97, sn99}.
Therefore,  changes in the slope of this relation to match observed reported values, which  are on average
 close to  $\alpha \sim 4$, might require the action
of other physical processes.
Theoretical  findings suggest that the triggering of powerful mass-loaded galactic 
winds by massive Supernova (SN) explosions would have 
an efficiency which depends on the virial halo so that larger effects are expected in low 
mass systems
\citep[][]{larson04}. In fact, \citet{ds86}  estimated analytically that
haloes with circular velocity smaller than $\sim 100$ km s$^{-1}$  would be strongly affected by massive SN events.
Nonetheless, the role of galactic outflows on the origin of the relations between the stellar
and dynamical properties of dwarf
galaxies still constitutes an open problem \citep{maclow99, tassis08}.
In a  cosmological framework, only recently it has been possible to start simulating  the triggering of galactic 
outflows on more physical basis  \citep[e.g.][]{ scan06, stin07}.
In fact, cosmological simulations that include this process obtained  sTFRs in more qualitative agreement with observations \citep{gov07, croft09}.
There is evidence 
that the implementation of a suitable feedback model could account for a bend in the TFR as suggested by
semi-analytical works
\citep[e.g.][]{kang05, nag05, vdbosh09, guo09} and some hydrodynamical simulations 
\citep[e.g.][]{tassis08}.
 However, a detailed analysis of the effects of SN feedback on the TFR using
a physically motivated model was still missing.

In this paper, we present results on the analysis of the sTFR and bTFR by using a 
SN feedback model specially suitable to
study the formation of galaxies in cosmological scenarios \citep{scan05, scan06} since it
does not include any scale-dependent parameter and yet, it is
capable to produce the expected anti-correlation between the strength of the galactic outflows and
the potential well of the haloes,  as shown by \citet{scan06}.

This paper is organized as follows. Section \ref{sec:simus} describes the numerical experiments. Section 
\ref{sec:results} outlines
the results. In Section \ref{sec:discussion} we discuss about our main findings and in Section 
\ref{sec:conclusions} we present the conclusions of this work.

\section{Numerical Experiments}

\label{sec:simus}

We performed cosmological simulations of a typical field region of the Universe 
consistent with the concordance model.  We adopt the following cosmological parameters: $\Omega_{\Lambda}=0.7$, $\Omega_{\rm m}=0.3$, $\Omega_{b}=0.04$, a normalization of the power 
spectrum of $\sigma_{8}=0.9$ and $H_{0}= 100 \ h \ {\rm km} \ {\rm s}^{-1}\ {\rm Mpc}^{-1}$, with $h=0.7$.
Two of the simulations have been ran with $2 \times 230^3$ (S230) particles while a third  and a fourth one used $2 \times 320^3$
(S320) and $2\times 160^3$ (S160) to check numerical 
effects. 
In all cases,  the simulated volumes correspond to a cubic box of a comoving
10 Mpc $h^{-1}$ side length.
The particle masses for the dark and initial gas components are given 
in Table \ref{tab:simus}.
Simulation S320 could only reach $z \approx 2$ due to high computational costs.
By analysing the sTFR and bTFR in S230, S320 and  S160, we  check our results for numerical effects. 

\begin{table}
\caption{\noindent Cosmological Hydrodynamical Simulations studied in this paper.}
\label{tab:simus}
{\center
\begin{tabular}{lcccc}\hline \hline
Name & F/NF  &  $N_{\rm p}$ & $M_{\rm dark}$ & $M_{\rm gas}  $  \\ 
     &       &              &$[10^6 {\rm M}_{\odot} h^{-1}]$ & $[10^6 {\rm M}_{\odot} h^{-1}]$ \\ \hline
S160 & F    & $2 \times160^3$  & $17.6$  &  $2.71$    \\ 
S230 & F   & $2 \times230^3$  & $5.93$  &  $0.91$    \\ 
S230NF & NF    & $2 \times230^3$  & $5.93$  &  $0.91$    \\ 
S320 & F   & $2 \times320^3$  & $2.20$  &  $0.34$    \\ 
\hline \hline
\end{tabular}
}
\\
Col. 1: Name of the simulation. 
Col. 2: Model with/without SN feedback (F/NF). 
Col. 3: Initial number of particles in the simulation. 
Col. 4: Mass of dark matter particles. 
Col. 5: Initial mass of gas particles.
\end{table}

The simulations have been performed by  using a version of {\small GADGET-3},
an update  of  {\small GADGET-2} optimized for massively parallel simulations of highly inhomogeneous systems \citep[][]{sh03, springel2005}. This version of {\small GADGET-3} includes the multiphase model for the interstellar medium (ISM) and the SN feedback scheme of \citet{scan05, scan06}.
  We run the same initial condition with (S230) and without  SN feedback (S230NF) 
in order to quantify the effects of SN-induced galactic winds.

The SN feedback scheme adopted in this work considers  both Type II and Type Ia SNe for the chemical and energy production. We adopt $0.7 \times 10^{51}$ erg per SN event.
The chemical evolution model used in this code is the one developed by  \citet{mos01} and, later on, 
 adapted  by \citet{scan05}  for {\small GADGET-2}. 
This model assumes the instantaneous recycling approximation (IRA) for SNII\footnote{\citet{scan05} also 
implemented a version that relax the IRA
assuming the age-mass-metallicity fitting polynomials estimated by \citep{raiteri96}. However, no significant changes were found in the results compared 
to the IRA.}.
 Lifetimes for the progenitors of
SNIa are randomly selected  in the range $[10^8, 10^9]$ yr. The chemical yields for SNII are given by \citet{WW95}
while those of SNIa correspond to the W7 model of \citet{thiel93}. 

The SN feedback model is grafted onto a multiphase model specially designed to improve the description  of the ISM, 
allowing the coexistence of diffuse and dense gas phases. Within this new framework, the injection of energy and heavy
elements into different components of the ISM can be more realistically treated leading to a more effective production
of galactic winds.  The ejected energy is distributed into  and managed  differently by  the so-called cold (temperature $T < T_c $ where $ T_c = 8\times 10^4$ K and density $\rho > 0.1 \rho_{\rm c}$ where $\rho_{\rm c} $ is $7 \times 10^{-26} {\rm g \ cm^{-3}}$)
and hot (otherwise) ISM components. The  hot phase thermalizes instantaneously the energy while the  
cold phase builds up a  reservoir until
it accumulates enough energy to raise the entropy of the gas particle to match the value  of 
 its own surrounding hot  environment.
 The properties of the hot and cold 
gas components of each gas particle are estimated 
locally without using any global properties of the systems.
The fraction of SN energy distributed into the cold phase is given by the feedback parameter $\epsilon_c$.
The ejection and distribution of chemical elements is coupled to the energy procedure so that the fraction
of metals pumped into the cold and hot phases are also given by $\epsilon_c$ in these simulations.
We adopt  $\epsilon_c =0.5$ which have been  found by \citet{scan08} to best reproduce a disc galaxy in simulations of a Milky-Way type halo.

Virialized structures are selected from the general mass distribution 
by using a standard friends-of-friends technique. The substructures residing
within a given virialized halo
are then individualized with the SUBFIND algorithm of \citet{springel01} to build up galaxy catalogues  at different 
redshifts.  
Note that, by definition, a substructure can correspond either to the central galaxy or a satellite
system within a given halo.
In order to characterize the simulated galaxies, 
we define the baryonic radius ($R_{\rm bar}$) as the one that encloses 83 per cent of the baryons associated
to each substructure. 
In the case of S160, S230 and S230NF, for all  the linear regressions, we only consider those simulated galaxies  defined by more than 
$2 \times 10^{3}$ particles within  $R_{\rm bar}$, which is equivalent to have stellar masses larger than $10^9$M$_\odot$ h$^{-1}$. In S230, Milky-Way type galaxies at $z = 0$ are resolved with approximately 
$10^{5}$ total particles within  $R_{\rm bar}$.  A similar procedure 
was carried out for the high resolution simulation S320, where Milky-Way type systems are already 
 resolved  with more than $10^{5}$ particles within  $R_{\rm bar}$ at $z = 2.0$. In S320, the smallest
simulated galaxies we considered for calculations have $10^{4}$  particles within $R_{\rm bar}$.

\begin{figure*}
\begin{center}\hspace*{-1.2cm}
\resizebox{4.5cm}{!}{\includegraphics{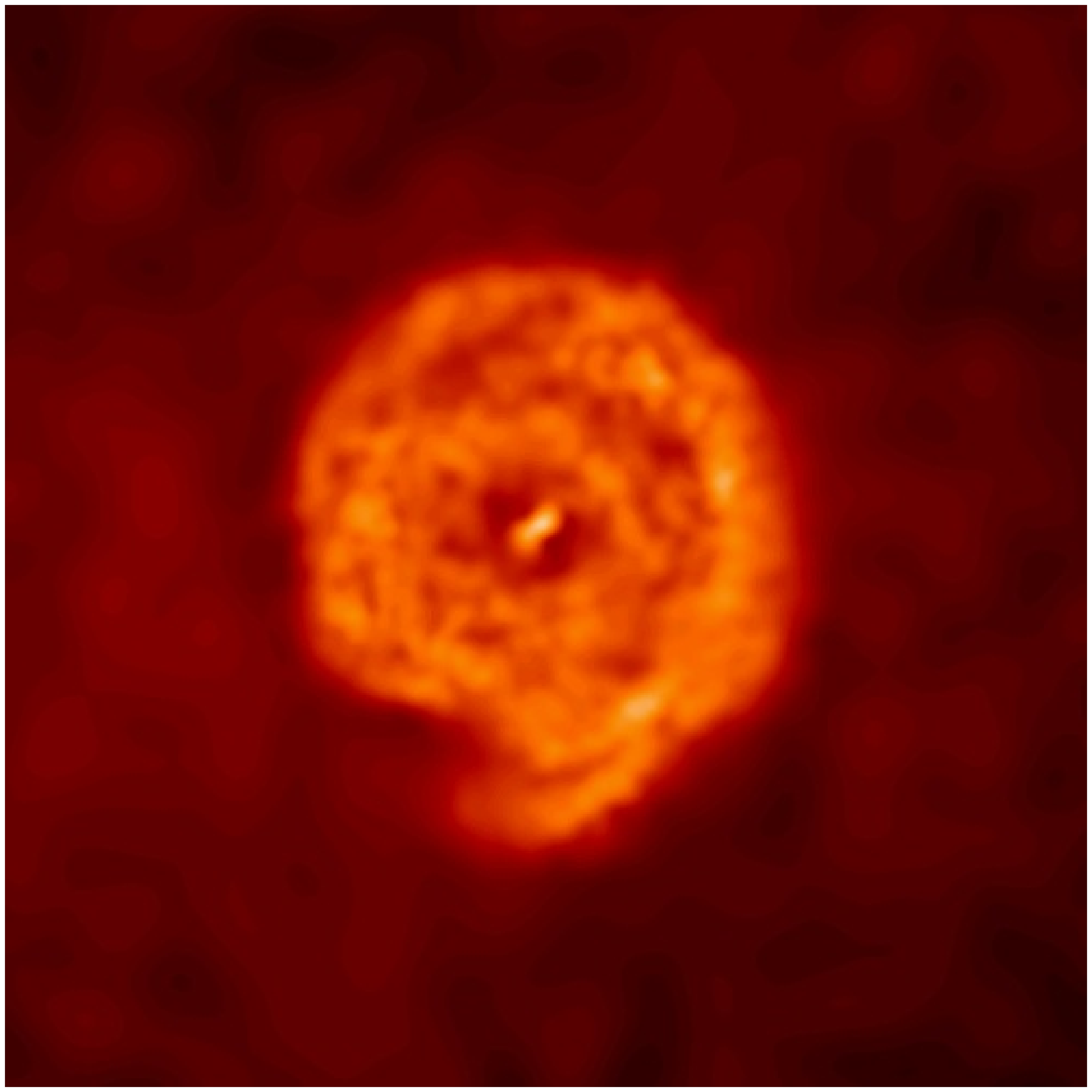}}
\resizebox{4.5cm}{!}{\includegraphics{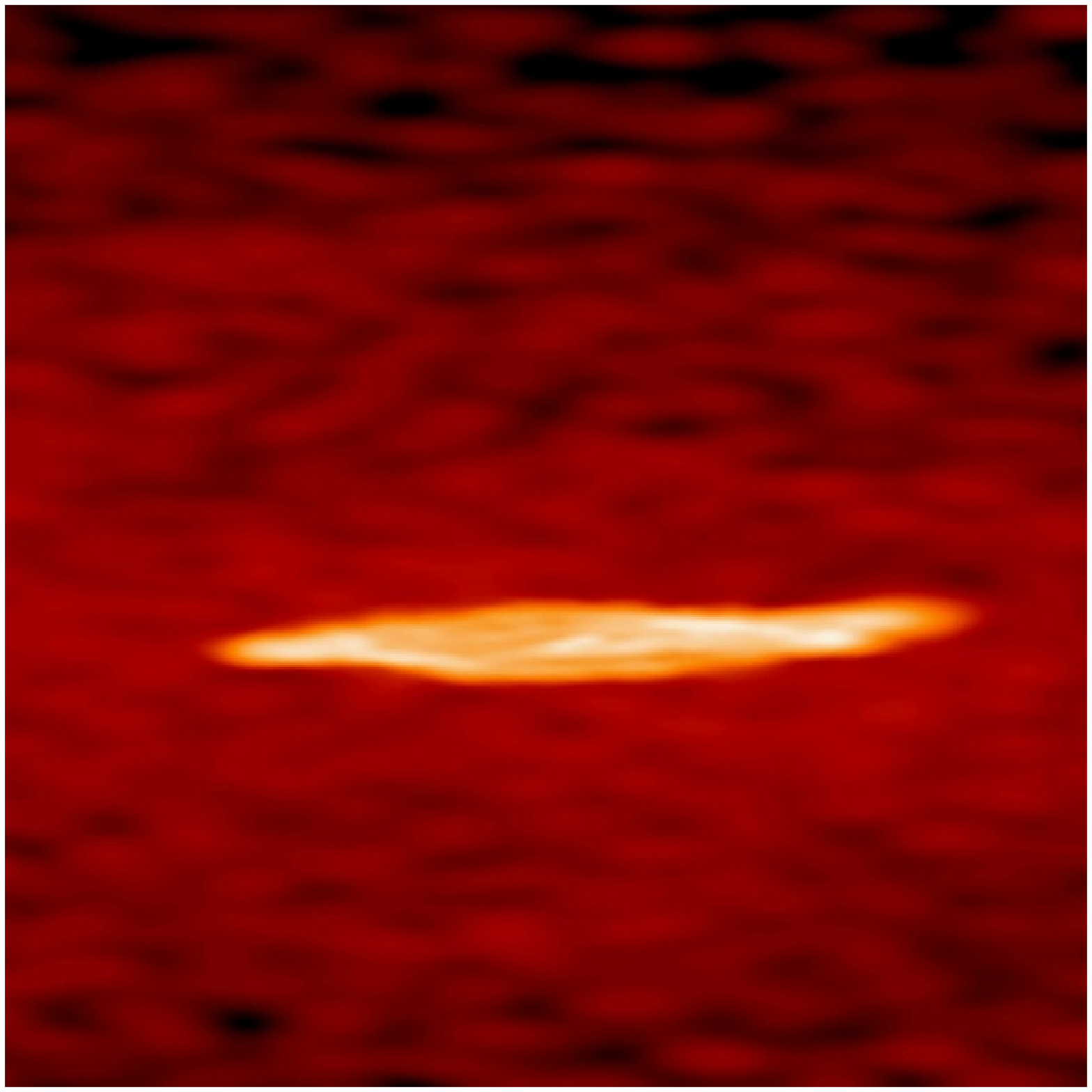}}
\resizebox{4.5cm}{!}{\includegraphics{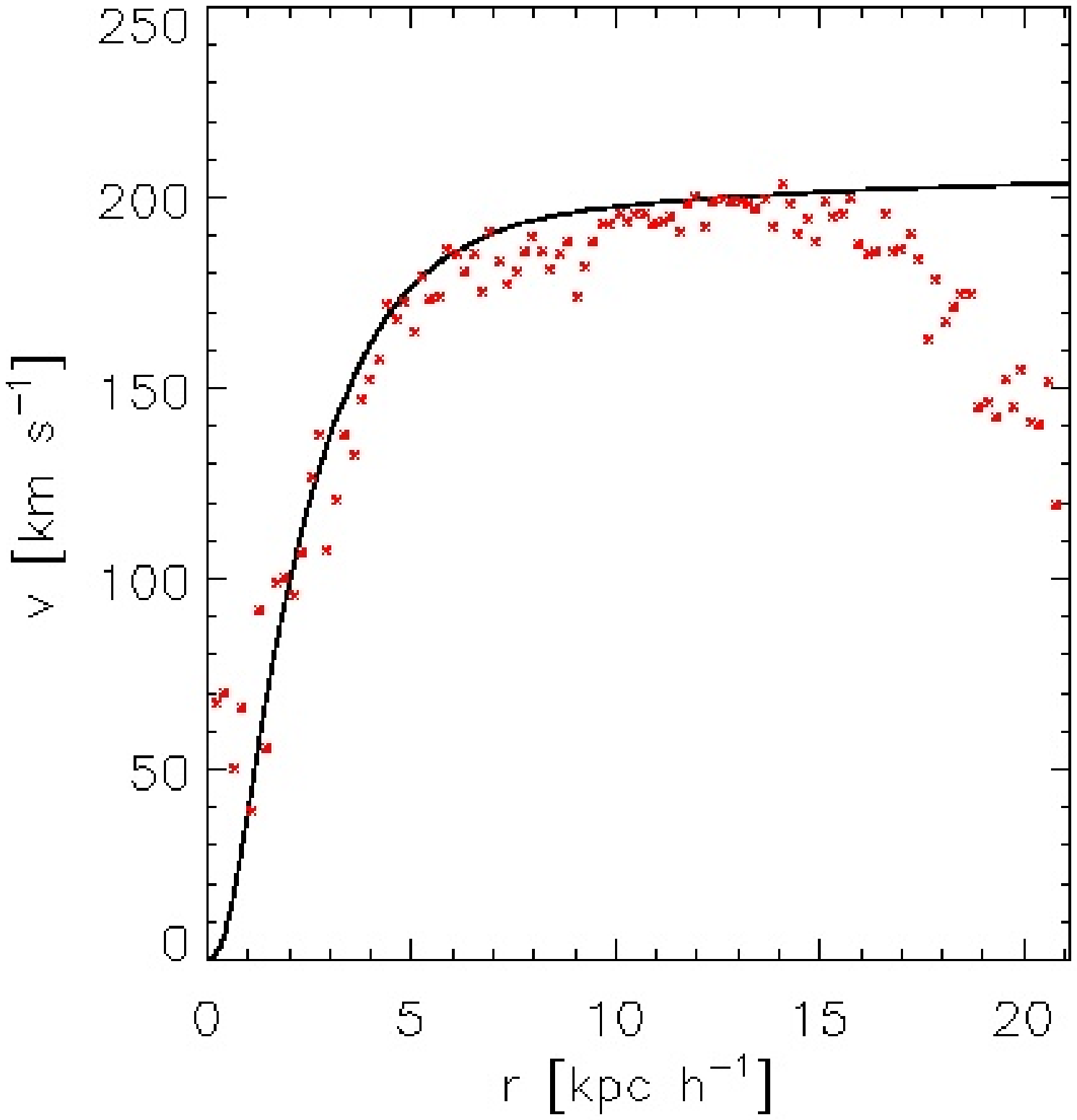}}
\end{center}
\caption[Rotation Curve]
{Projected gaseous mass distributions for a typical disc galaxy in a face-on (left panel) and an edge-on (middle panel) views and the corresponding $V_{\rm rot}$ (dots)  and $V_{\rm circ}$ (continuous line) for this system (right panel) in S230.
For this galaxy, $R_{\rm bar}$  is $\approx 13.7$ kpc h$^{-1}$. }
\label{fig:rot_curv}
\end{figure*}

The properties of the simulated galaxies such as   stellar mass, baryonic mass and rotation curves 
are estimated within $R_{\rm bar}$. 
We define as disc systems those which have more than 75 per cent of their gas component on a rotationally supported
disc structure by using the condition  $\sigma/V 
< 1$  to select them (where $\sigma$ and $V$ are the velocity dispersion and the tangential velocity component, respectively).
We acknowledge the fact that, at $z=0$, most of our galaxies contains large stellar bulges and thick stellar discs,
 being more consistent
with early type spirals.
 Nevertheless, the gaseous disc components are very well-defined and 
 trace remarkably well the potential well of their host haloes as   can clearly be appreciated in Fig. \ref{fig:rot_curv}.
Rotation curves are estimated by using the tangential velocity of gas particles on the plane 
perpendicular to their total angular momentum. For these systems, tangential velocity 
constitutes a good representation of the potential well of the system. For the sake of simplicity, we use $V_{\rm cir}^2 = G M(r< R)/R$  (where $M(r< R)$ is the enclosed total mass within $R$) estimated at $R_{\rm bar}$ as an indicator of the gravitational mass for systems in both S230 and S230NF.

\begin{figure*}
\begin{center}
\resizebox{7cm}{!}{\includegraphics{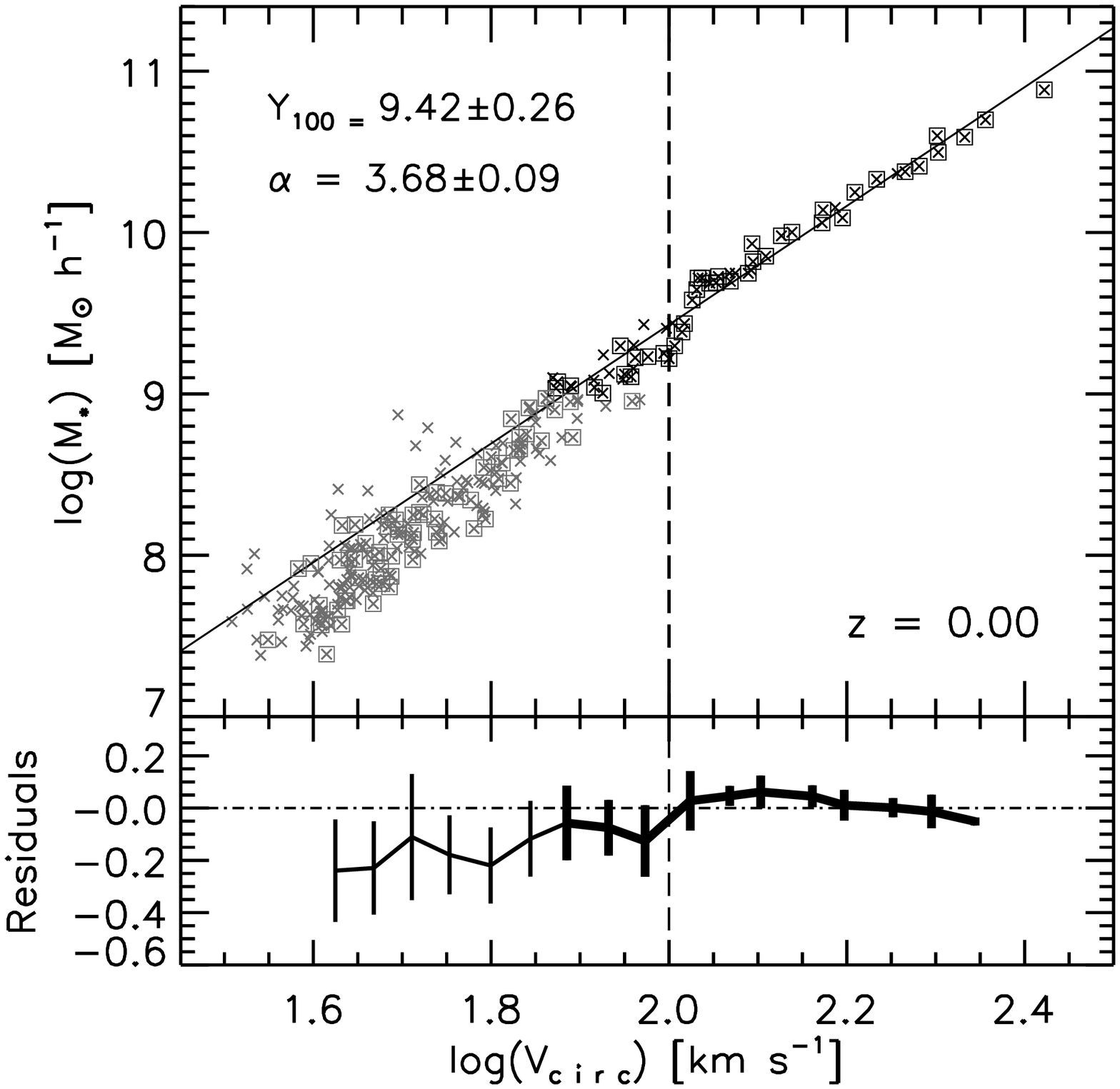}}
\resizebox{7cm}{!}{\includegraphics{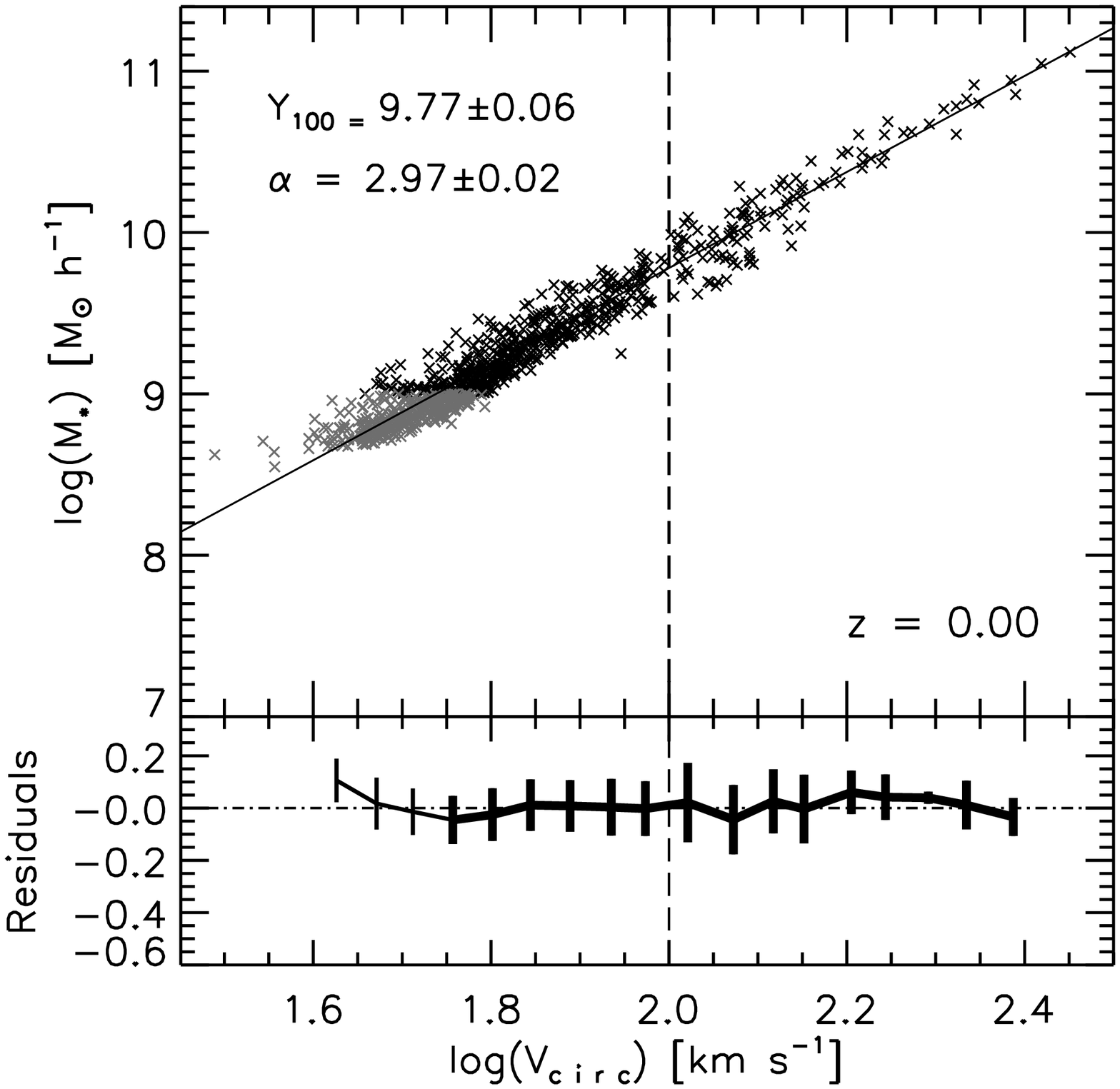}}\\
\resizebox{7cm}{!}{\includegraphics{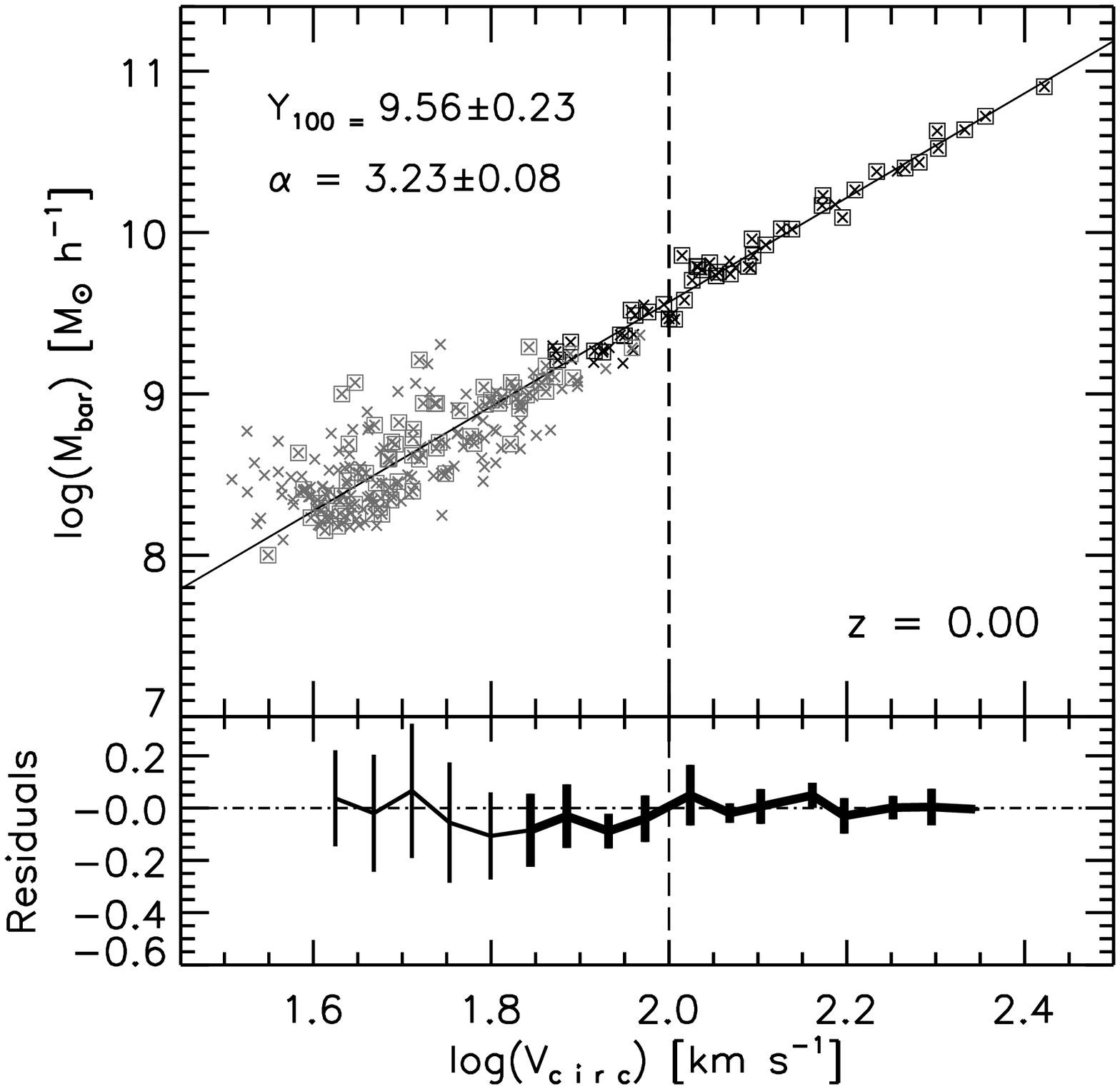}}
\resizebox{7cm}{!}{\includegraphics{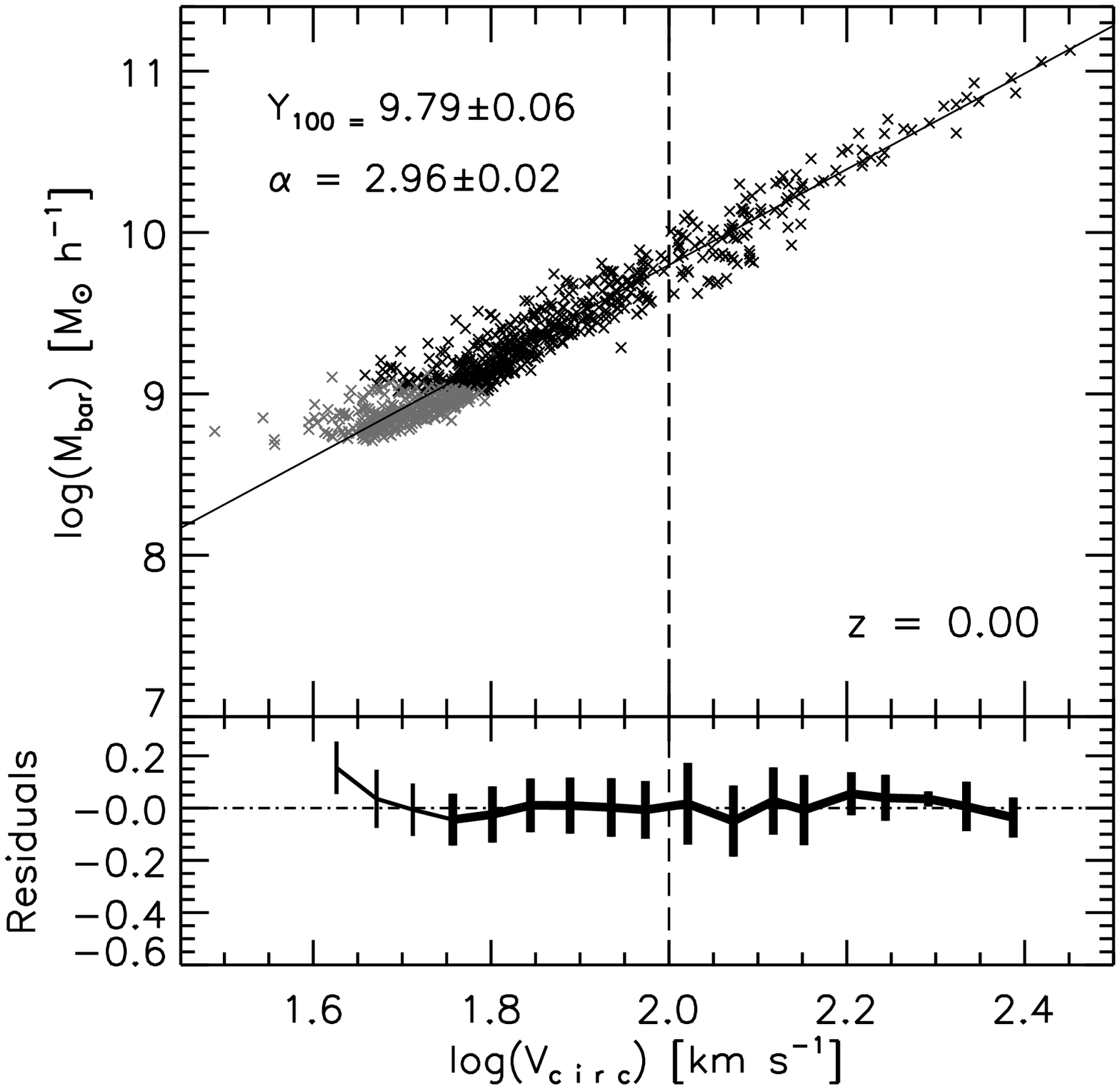}}
\end{center}
\hspace*{-0.2cm}
\caption{The sTFR (upper panels) and bTFR (lower panels) in the  S230 (left) 
and S230NF (right) runs, at $z = 0$  .
Grey symbols depict low mass systems ($M_{*} <  10^9  {\rm M_{\odot}} h^{-1}$),
while massive ones ($M_{*} >  10^9  {\rm M_{\odot}} h^{-1}$)
are coloured in black.
Systems in S230 run, with well-defined gaseous discs are
enclosed with squares. The black line corresponds to the linear fit.  The values
of the slope ($\alpha$) and $Y_{100}$ are indicated in the figure.
The vertical lines indicates the velocity where the residuals (small boxes) of the sTFR in S230 depart systematically from zero.
The thicker lines denote the range of velocities associated with stellar masses larger than $10^9  {\rm M_{\odot}} h^{-1}$.
} 
\label{tf}
\end{figure*}

\section{Results}

\label{sec:results}

We calculated the sTFR for our simulated galaxies by the fitting of a relation of the form 
$\log M_{\rm *} = \alpha \log (V_{\rm cir}/100 \ {\rm km \ s}^{-1}) + Y_{100}$.
In the case of the local sTFR in S230, we found  a slope of $3.68 \pm 0.09$ and $Y_{100}= 9.42 \pm 0.26$ in general agreement
 with observations \citep{mcg00, bdj01}.
As  can be   seen  from Fig.~\ref{tf} (upper left panel), the residuals of the linear fit to the relation (small box) depart systematically from zero at around 100  km s$^{-1}$,  so that systems with lower circular velocities tend to 
have smaller stellar
masses than the predictions obtained from the fitting. Interestingly, this trend is consistent with the observations  reported by \citet{mcg00},
and from theoretical expectations \citep{larson04,ds86}. In order to analyse at which extent SN feedback might be responsible for this behaviour, we compared these findings with the ones obtained for S230NF 
as shown in the upper right panel of Fig.~\ref{tf}. We can appreciate that SN feedback seems to be crucial to reproduce the observed features in the sTFR since when this mechanism is turned off,  the simulated sTFR exhibits a linear behaviour
with a slope of $ \sim 3$ in agreement with theoretical predictions.
We can also see that the wind-free run predicts larger stellar masses for systems at a given circular 
velocity since it cannot regulate the transformation of gas into stars which is too efficient in hydrodynamical simulations.

We also studied the bTFR in  S230 and S230NF (Fig.~\ref{tf}, lower panels). The bTFR is obtained by adding
all baryons within $R_{\rm bar}$, regardless of its physical state.
We can see that both models (with and without SN feedback) predict a linear
trend for the bTFR, at least, for the range of velocities covered by these simulations.
At $z=0$, the simulated bTFR in S230 has a slope of $3.23 \pm 0.08$ and  $Y_{100} = 9.56 \pm 0.23$. 
These values are in general good  agreement with observational results reported for late and early type
 galaxies \citep{rijcke07, guro10}. However, we warn that the comparison with observations is tricky since
we sum up the total amount of gas, while observers are limited by the instrumental techniques
which give only access to gas mass with certain physical properties.
In the case of S230NF, the bTFR has a larger value of $Y_{100}$ because these galaxies have been able to
retain their baryons in the central regions since no SN-driven outflows could be triggered.  We also note that S230NF 
yields similar fittings to the sTFR and bTFR, indicating that stars dominate
the baryonic phase at this redshift.

\begin{figure*}\hspace*{0.5cm}\\
\hspace{1cm}
\resizebox{7cm}{!}{\includegraphics{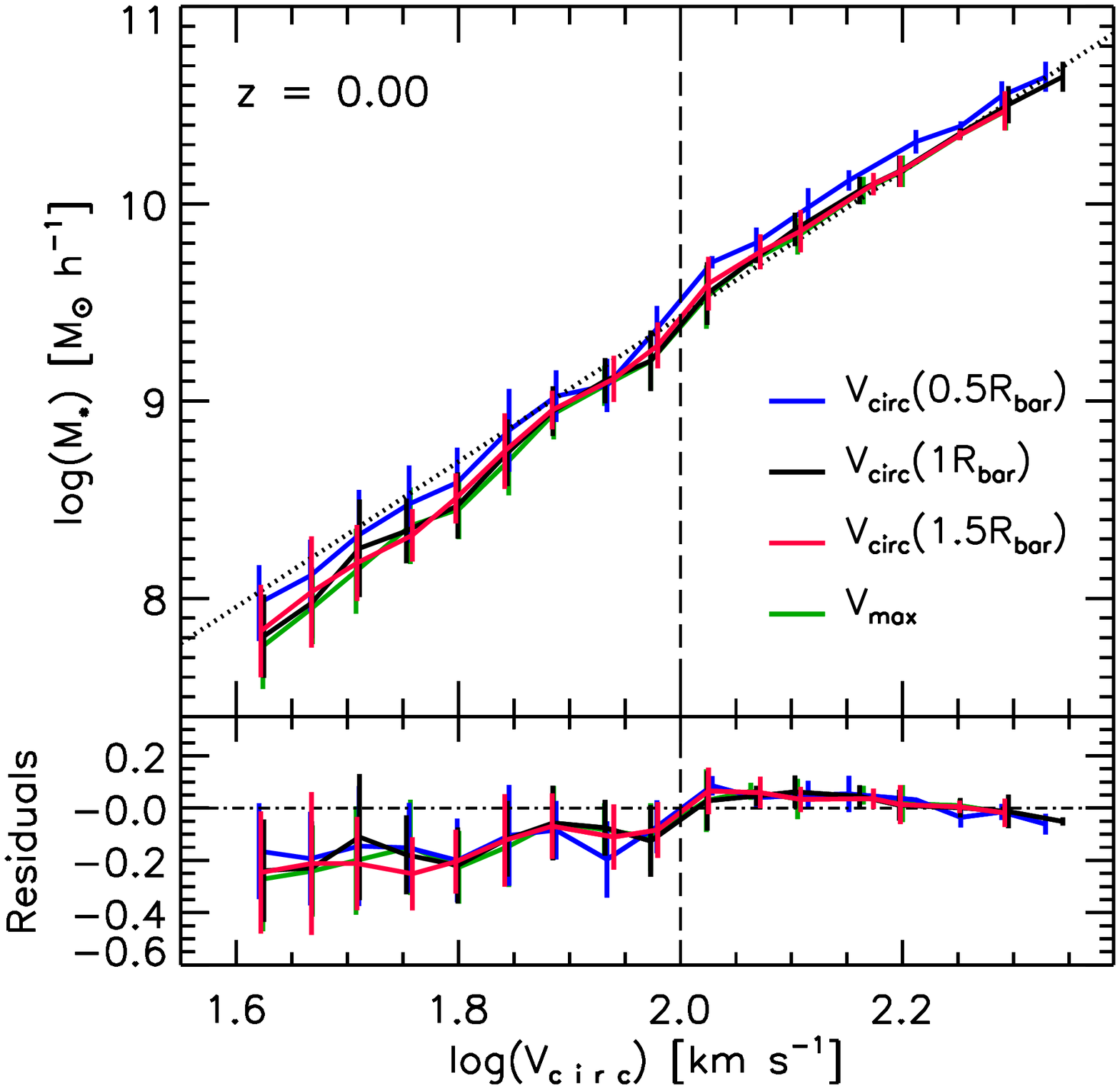}}
\hspace{1cm}
\resizebox{7cm}{!}{\includegraphics{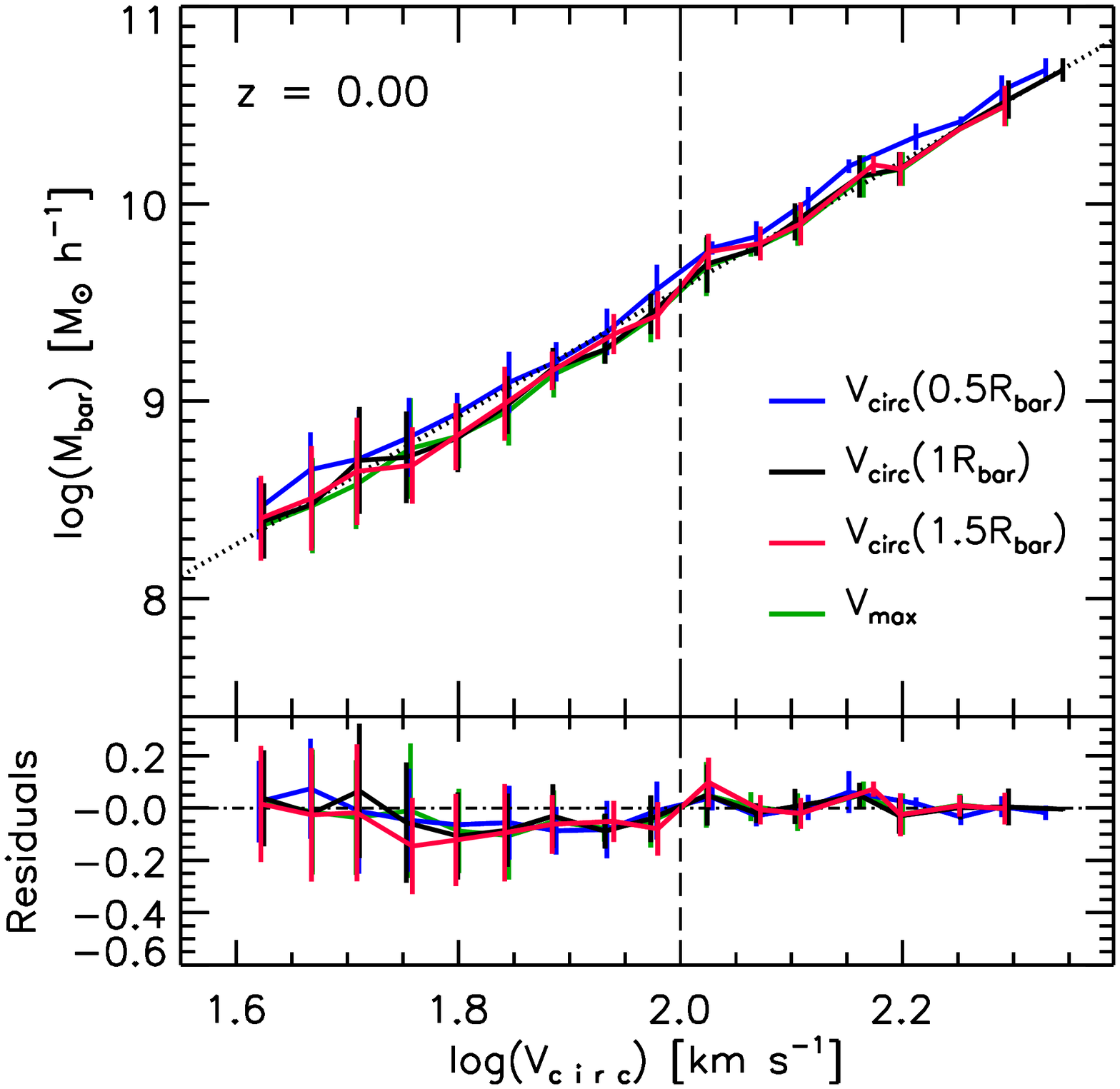}}
\caption{
Mean sTFR (left panel) and bTFR (right panel), and the corresponding standard
deviations for S230 at $z=0$.  Results obtained by using
different kinematic indicators are compared: $V_{\rm circ}$ at $0.5 R_{\rm bar}$ (blue),
$R_{\rm bar}$ (black) and $1.5 R_{\rm bar}$ (red), and the maximum value
of the rotation curve $V_{\rm max}$ (green).
The dotted black lines represent the fittings to the  high mass-end 
($M_{*} >  10^9  {\rm M_{\odot}} h^{-1}$) of the simulated TFRs when using
$V_{\rm circ}$ at $R_{\rm bar}$ as kinematic indicator.
The vertical line depicts the characteristic  velocity where the sTFR bends.
The small boxes show the residuals of each TFR with respect to the linear fit to its high-mass end.
}
\label{velos}
\end{figure*}

As we have already mentioned, the previous analysis has been done by
using the value of $V_{\rm circ}$ at $R_{\rm bar}$ as a kinematical indicator.  
In order to analyse the dependance of our results on the particular choice of the radius
at which we measure $V_{\rm circ}$, we also compared the relations obtained by using $V_{\rm circ}$
evaluated at $0.5 R_{\rm bar}$ and $1.5 R_{\rm bar}$.
We also estimated the maximum value of the rotation curve $V_{\rm max}$ as can be seen
from Fig.~\ref{velos}.
The simulated sTFR and bTFR are not
significantly affected by variations in the radius at which $V_{\rm circ}$
is estimated.  The largest changes with respect to our previuos results (Fig.~\ref{tf}) 
are obtained when using $V_{\rm circ}$ at $0.5 R_{\rm bar}$ as the velocity estimator. 
Nevertheless, all changes remain whithin
a ${\sigma}$.
It is also worth noting that, in these simulations,  $V_{\rm circ}$ at $R_{\rm bar}$ constitutes
a good proxy for the velocity $V_{\rm max}$, which is commonly employed in many
observational works. 
Tables \ref{tab:fits_stfr} and \ref{tab:fits_btfr} summarize the 
parameters corresponding to the linear 
fits to the high-mass end ($M_{*} >  10^9  {\rm M_{\odot}} h^{-1}$)  of the TFRs shown in Fig.~\ref{velos}.
We can appreciate that, in all cases, the values of $\alpha$  and $Y_{100}$ agree within a ${\sigma}$.
Regarding the characteristic velocity where the sTFR bends, our findings suggest that
it does not depend on the particular kinematical estimators tested in this paper (Fig.~\ref{velos}, small boxes).
In the light of these results, hereafter  we will continue using 
$V_{\rm circ}$ at $R_{\rm bar}$ as the kinematical indicator for our calculations.

\begin{table}
\caption{\noindent Results of the linear fits 
to the high-mass end ($M_{*} >  10^9  {\rm M_{\odot}} h^{-1}$) of the sTFR  
for S230 at $z=0$.}
\label{tab:fits_stfr}
{\center
\begin{tabular}{lcc}\hline \hline
$V_{\rm i}$ & $Y_{100}$  &  $\alpha$ \\  \hline
$V_{\rm circ}$ at $0.5 R_{\rm bar}$  & $9.51 \pm 0.26$    & $3.62 \pm 0.09$    \\
$V_{\rm circ}$ at $R_{\rm bar}$      & $9.42 \pm 0.26$    & $3.68 \pm 0.09$    \\
$V_{\rm circ}$ at $1.5 R_{\rm bar}$  & $9.43 \pm 0.25$    & $3.59 \pm 0.08$    \\
$V_{\rm max}$                        & $9.41 \pm 0.25$    & $3.67 \pm 0.08$  \\ \hline 
\hline \hline
\end{tabular}
}
\\
Col. 1: Kinematic indicator used for the calculations.
Col. 2: Stellar mass corresponding to a velocity of $100 {\rm \ km \ s}^{-1}$ according to the fitting. 
Col. 3: Slope derived from the linear regression.
\end{table}

\begin{table}
\caption{\noindent Results of the linear fits 
to the high-mass end ($M_{*} >  10^9  {\rm M_{\odot}} h^{-1}$) of the bTFR 
for S230 at $z=0$.}
\label{tab:fits_btfr}
{\center
\begin{tabular}{lcc}\hline \hline
$V_{\rm i}$ & $Y_{100}$  &  $\alpha$ \\  \hline
 $V_{\rm circ}$ at $0.5 R_{\rm bar}$  & $9.64 \pm 0.20$    & $3.19 \pm 0.07$    \\
 $V_{\rm circ}$ at $R_{\rm bar}$      & $9.56 \pm 0.23$    & $3.23 \pm 0.08$    \\
 $V_{\rm circ}$ at $1.5 R_{\rm bar}$  & $9.57 \pm 0.25$    & $3.14 \pm 0.08$    \\
 $V_{\rm max}$                        & $9.55 \pm 0.24$    & $3.22 \pm 0.08$ \\ 
\hline \hline
\end{tabular}
}
\\
Col. 1: Kinematic indicator used for the calculations.
Col. 2: Baryonic mass corresponding to a velocity of $100 {\rm \ km \ s}^{-1}$  according to the fitting. 
Col. 3: Slope derived from the linear regression.
\end{table}

\begin{figure}\hspace*{0.5cm}\\
\begin{center}
\resizebox{7cm}{!}{\includegraphics{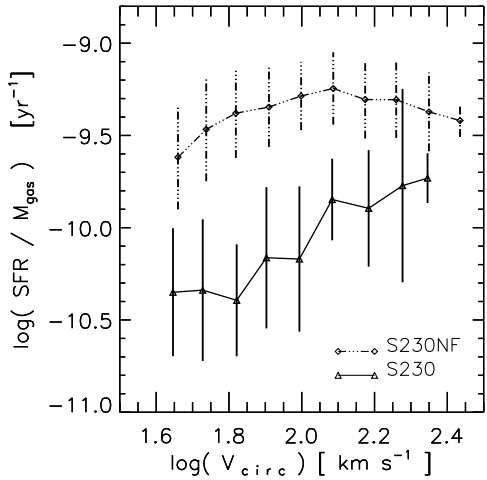}}
\end{center}
\caption{
Mean star formation efficiency (eSFR) as a function of $V_{\rm circ}$  for galaxies in  S230 (triangles)
and in  S230NF (diamonds) at $z=0$.
}
\label{fig:esfr}
\end{figure}

By comparing the sTFR and bTFR in S230, it is clear that adding the  gas
mass in the calculations contributes to restore the linearity of the TFR over a larger velocity range.
These findings
suggest that, at $z \sim 0$, the bend of the local sTFR 
might be partially caused by a  decrease of the star formation
rate in smaller systems as a consequence of the gas heating  by SNe.
To quantify at which extent SN feedback can generate a decrease in the star formation
activity of simulated galaxies, we calculated the star formation efficiency (eSFR)  defined as
the ratio of the star formation rate of a given galaxy to the total amount of gas contained within it.
In Fig. \ref{fig:esfr}, we compare the mean eSFR as a function of $V_{\rm circ}$ for galaxies in S230 and S230NF at $z=0$.
As expected, the model which includes SN feedback predicts smaller eSFRs at a given
$V_{\rm circ}$, with slow-rotating systems exhibiting the most important variations, on average. These systems still have an important fraction of gas but it is in the form of a  diffuse warm environment which does not fulfil the conditions to form stars.

\begin{figure}\hspace*{0.5cm}\\
\resizebox{8cm}{!}{\includegraphics{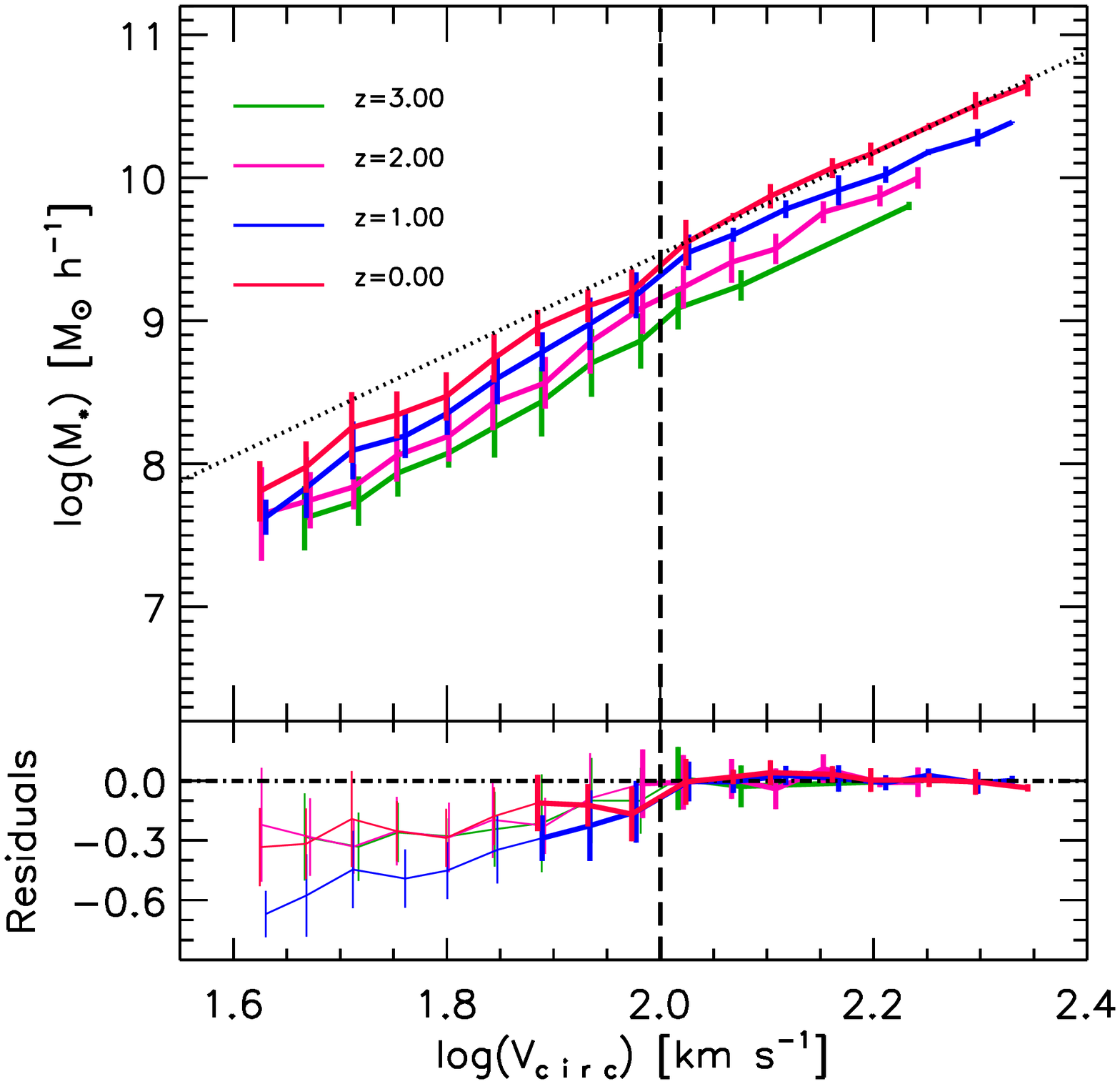}}
\resizebox{8cm}{!}{\includegraphics{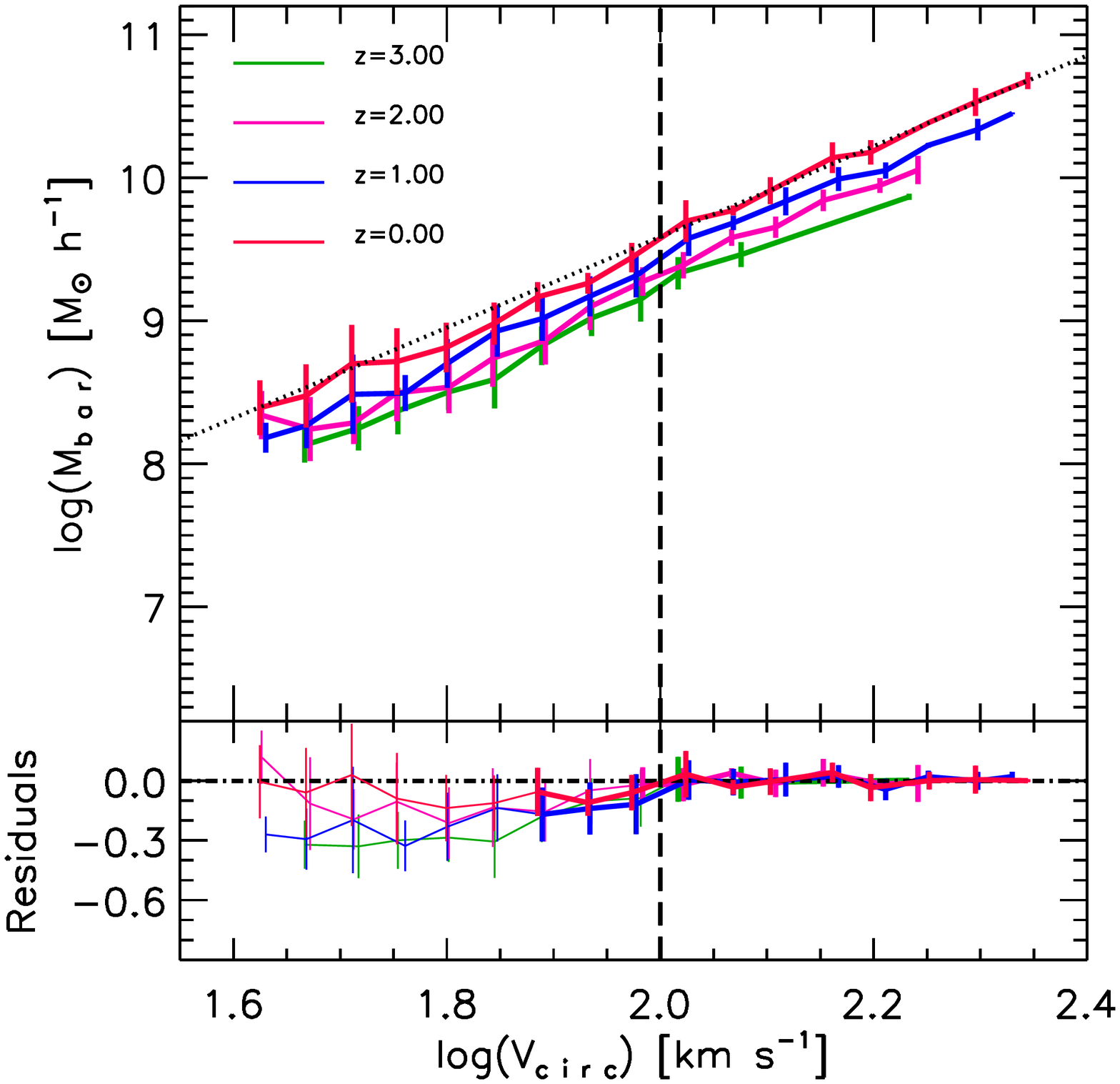}}
\caption{
The evolution of the mean sTFR (upper panel) and bTFR (lower panel) as a function of redshift 
for S230 for $z=3,2,1$ and $0$ (green, magenta, blue and red). The dotted black  lines correspond to the fittings to the  simulated TFRs for systems with 
 $V_{\rm circ} > 100 \ {\rm km \ s}^{-1}$ at $z=0$. 
The vertical line indicates this characteristic  velocity.
In the small boxes we included the residuals estimated with respect to
the corresponding linear regressions calculated at each redshift.
The thicker lines in the residuals indicate the velocities associated with 
stellar masses larger that $10^9  {\rm M_{\odot}} h^{-1}$.
}
\label{evol}
\end{figure}

To analyse the evolution of both the sTFR and bTFR as a function of redshift,
we performed linear fits to the fast rotators  using the characteristic  velocity $V_{\rm circ} =  100 \ {\rm km \ s}^{-1}$  as a reference value at each  analysed redshift.
 As  can be seen in Fig. \ref{evol},  the residuals of the sTFR show a systematic departure from zero  already from  $z \approx 3$ which  occurs at approximately a  similar characteristic velocity of $ \sim 100 \ {\rm km \ s}^{-1}$.
For this SN  model, we detect an evolution in the $Y_{100}$ of the sTFR of $\approx 0.44 $ dex since $z \approx 3$, while the
 slope of the fast rotators systems
remains almost constant.
In the case of the bTFR, $Y_{100}$ evolves by  around $\sim 0.30$ dex between $z=3$ and $z=0$.
In particular, as  can be appreciated from Fig. \ref{evol}, the simulated bTFR retains  the linear relation over  a larger velocity range. At $z =0$,  there is no clear signal for a change of the slope within the numerical dispersion.
This finding is not in disagreement with the  recent observational results of  \citet{mcg10} who reported  a break in the  bTFR at  $\sim 20$ km s$^{-1}$, because our simulations do not resolve numerically  such low velocity systems. 
While larger statistics and higher
numerical resolution is needed to test  the behaviour of the  bTFR for low velocity
systems, more detailed observations are also required to robustly probe the existence of the bend in the bTFR. 
With regard to the S230NF, both the sTFR and the bTFR exhibit a linear trend since $z=3$ with no significant changes
in the slope but with  an evolution of $\approx 0.55$ 
in $Y_{100}$, which is consistent with  the higher eSFR measured in  the galaxies in this run.
Both TFRs show larger dispersion for low velocity rotators which are also more
gas rich than systems at the high velocity end. 
At a given velocity,
slow rotators can have different baryonic and stellar masses evidencing their
different evolutionary paths. This behaviour can be also seen in S230NF albeit weaker.
Hence, the dispersion has at least two causes: the different evolutionary paths
at a given circular velocity which regulate the transformation of gas into stars and
the action of SN feedback.

\begin{figure}\hspace*{0.5cm}\\
\resizebox{8cm}{!}{\includegraphics{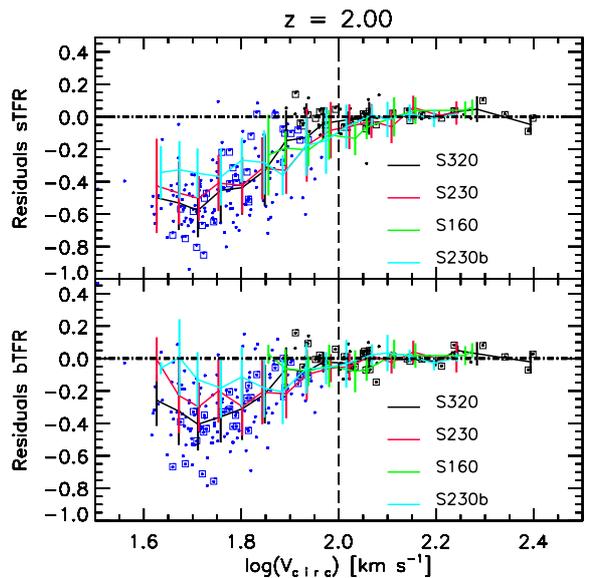}}
\caption{
Residuals corresponding to the
linear fits to the massive end ($M_{*} >  10^9  {\rm M_{\odot}} h^{-1}$) of the
sTFR (upper panel) and bTFR (lower panel) in S320 (symbols) at $ z = 2.0$.
Blue symbols depict low mass systems ($M_{*} <  10^9  {\rm M_{\odot}} h^{-1}$),
while massive ones ($M_{*} >  10^9  {\rm M_{\odot}} h^{-1}$)
are coloured in black.
Systems with well-defined gaseous discs are
enclosed with squares. 
The solid lines depict the mean values and standard deviations associated to the residuals of the linear fits of S320 (black), S230 (red), S160 (green) and S230b (cyan). 
}
\label{resol}
\end{figure}

To  assess the effects of numerical resolution on the determination of 
characteristic velocity $V_{\rm c} \sim 100 \ {\rm km \ s}^{-1}$, 
we performed linear fits to the massive end ($M_{*} >  10^9  {\rm M_{\odot}} h^{-1}$) of the 
sTFR and bTFR  in S320, S230 and S160  at $ z = 2.0$.
In Fig.~\ref{resol}, we compare the mean values and standard
 deviations of the residuals associated to
each simulation.  We also show with symbols the results for S320.  
As can be seen, the trend for the sTFR is clearly present in all simulations, showing a change in the slope at a similar velocity.
With respect to the bTFR, it also exhibits a bend in the residuals at around $ \sim 100 \ {\rm km \ s}^{-1}$ at $z \sim 2$ but the
trend is weaker than that of the sTFR.
The three simulations agree on  the existence of the departure from linearity at
approximately the same velocity. 
In the three runs, we  also find  larger dispersions  for low velocity rotators.
The agreement between the three different resolution-level simulations 
  suggests that these results are robust against numerical artefacts.

Finally, since the SN feedback model uses 
$T_c = 8\times 10^4$ K (equivalent to $\log V_{\rm vir} \approx 1.95$, where
$V_{\rm vir}$ is the virial velocity of the halo)
as a parameter to separate cold and hot gas-phases surrounding
stellar populations, we evaluated at which extent this selection can influence the
determination of the characteristic velocity where the sTFR bends.
To assess this issue, we performed an extra simulation (S230b) with the same
initial conditions and SN and star formation parameters of S230, except for $T_c$ which
was lowered to $5 \times 10^4$ K (equivalent to $\log V_{\rm vir} \approx 1.84$).
Fig.~\ref{resol} shows that for both TFRs the residuals of these  simulations are very similar. Particularly, both simulations predict
a bend at $ \sim 100 \ {\rm km \ s}^{-1}$ independently of the adopted value of 
$T_c$ indicating that this parameter has no significant effect
of the determination of the bend.

\section{Discussion}

\label{sec:discussion}

\begin{figure*}\hspace*{0.5cm}\\
\begin{center}
\resizebox{6cm}{!}{\includegraphics{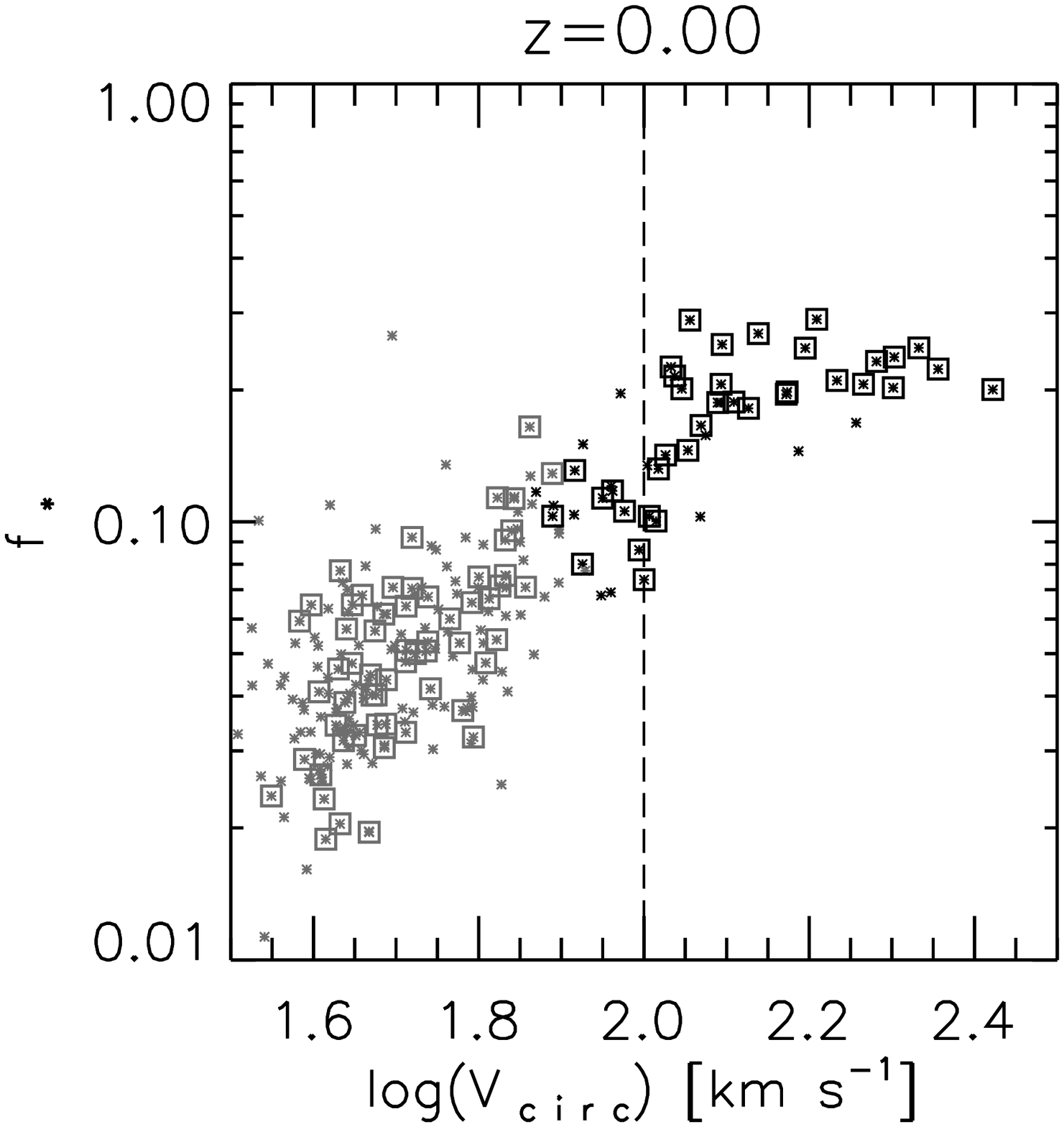}}
\resizebox{6cm}{!}{\includegraphics{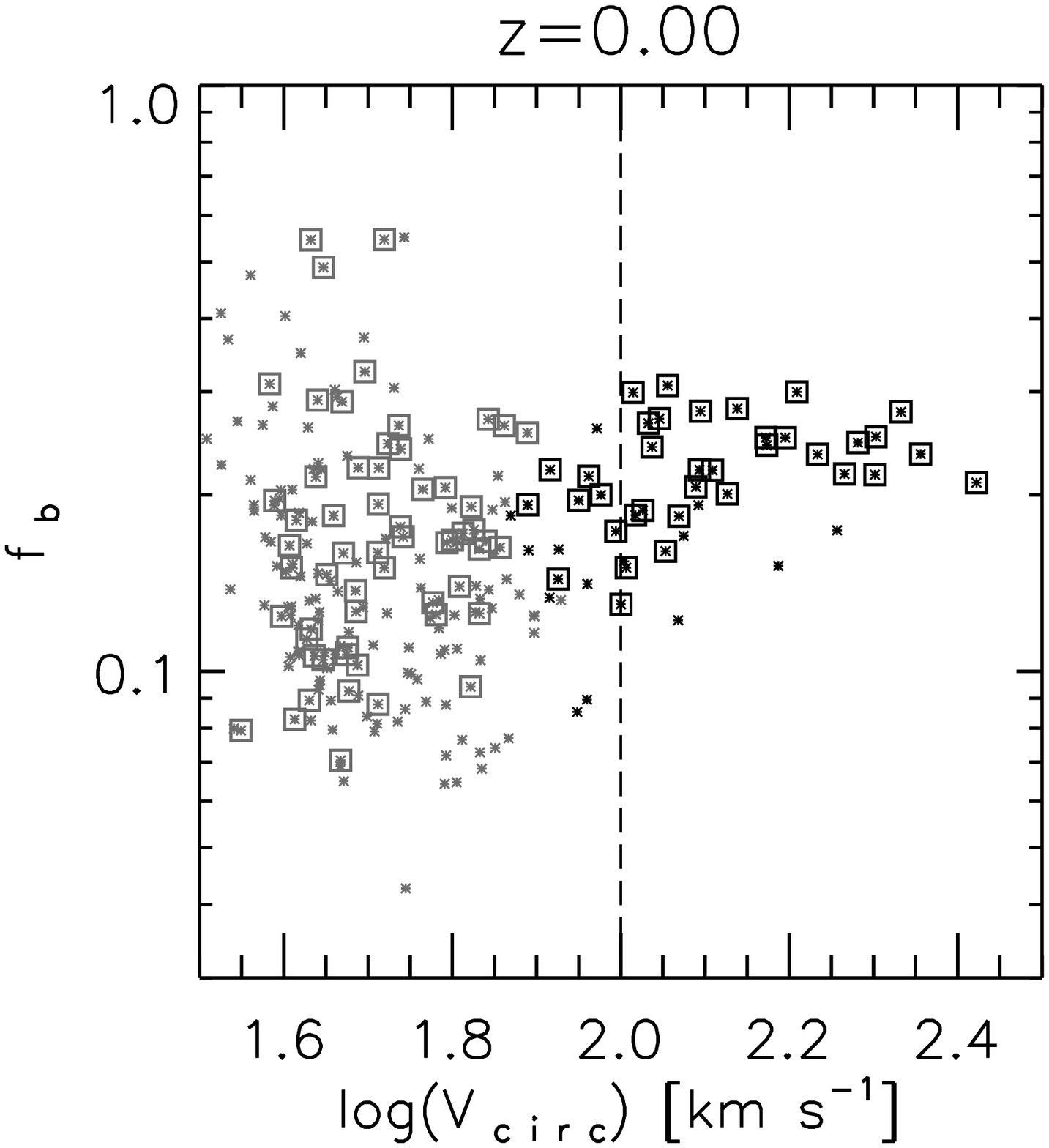}}
\resizebox{6cm}{!}{\includegraphics{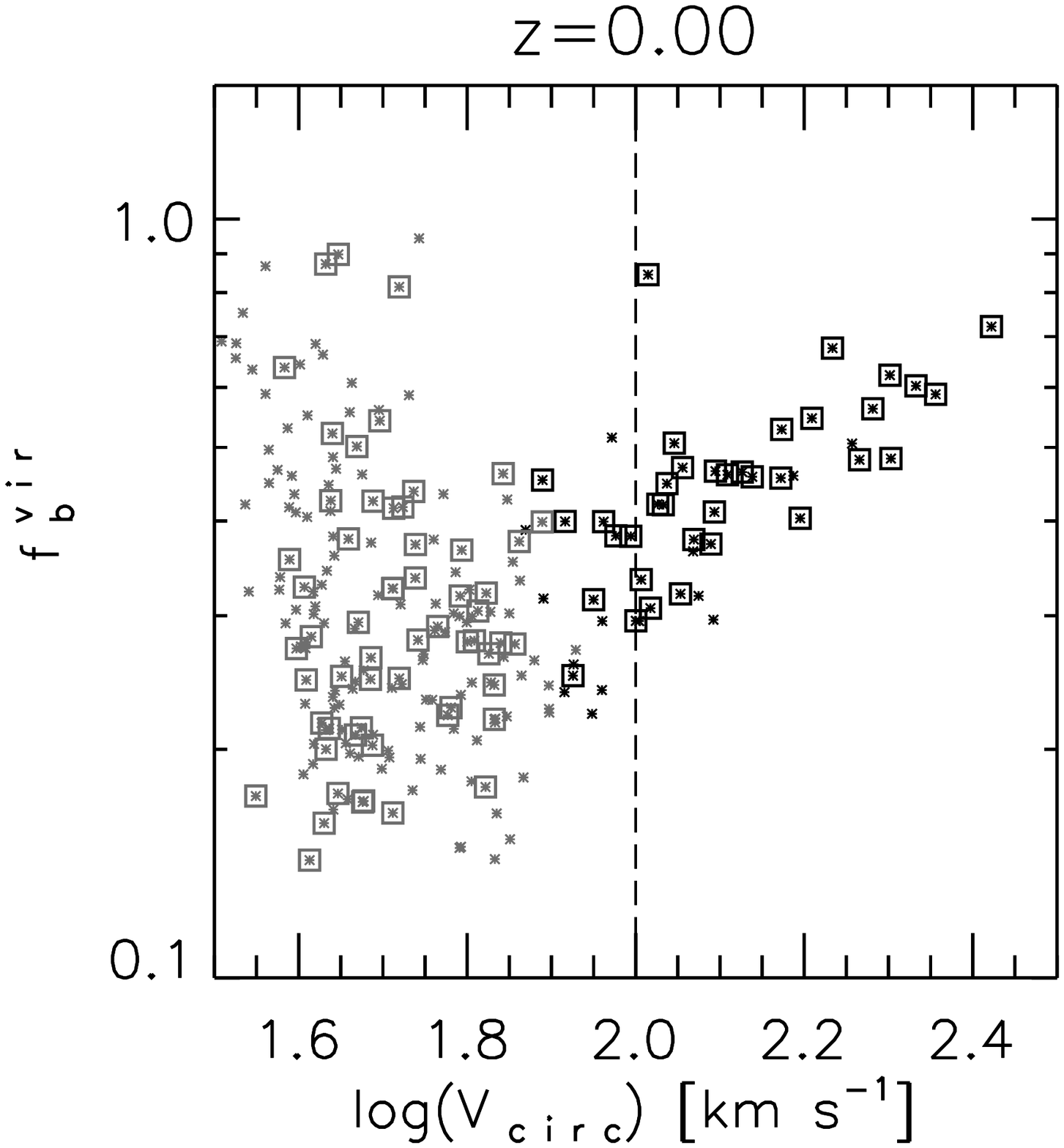}}\\
\resizebox{6cm}{!}{\includegraphics{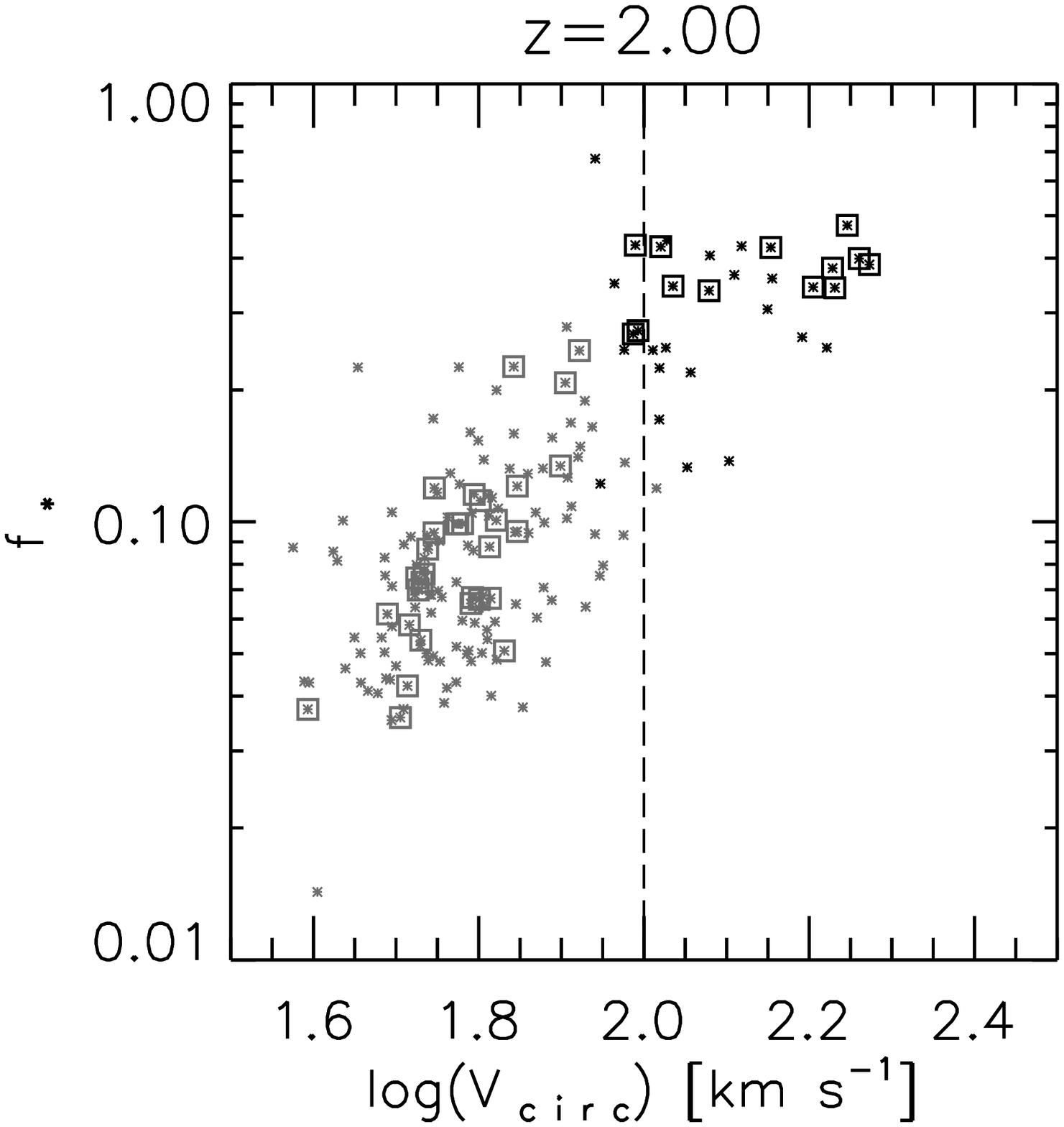}}
\resizebox{6cm}{!}{\includegraphics{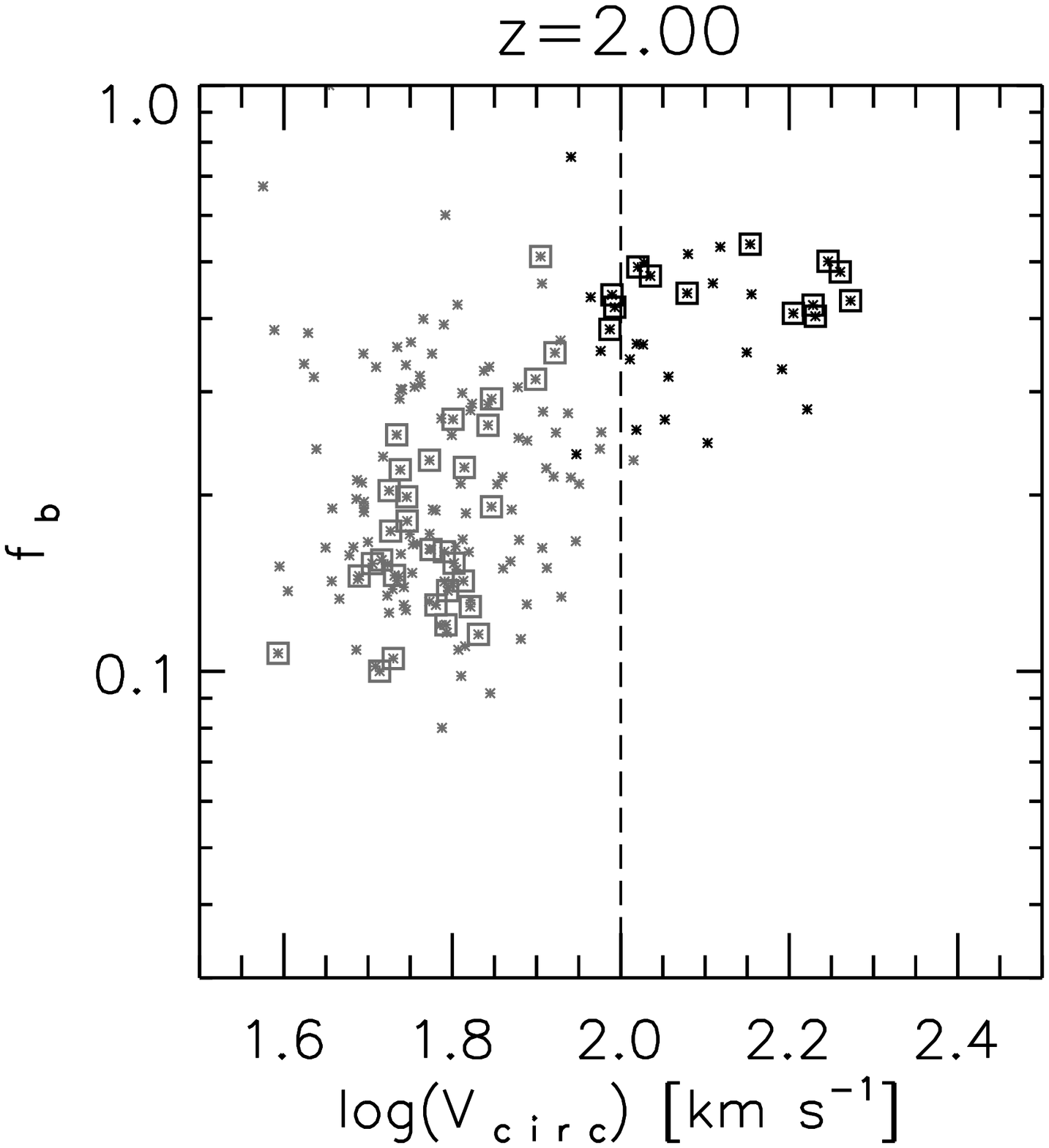}}
\resizebox{6cm}{!}{\includegraphics{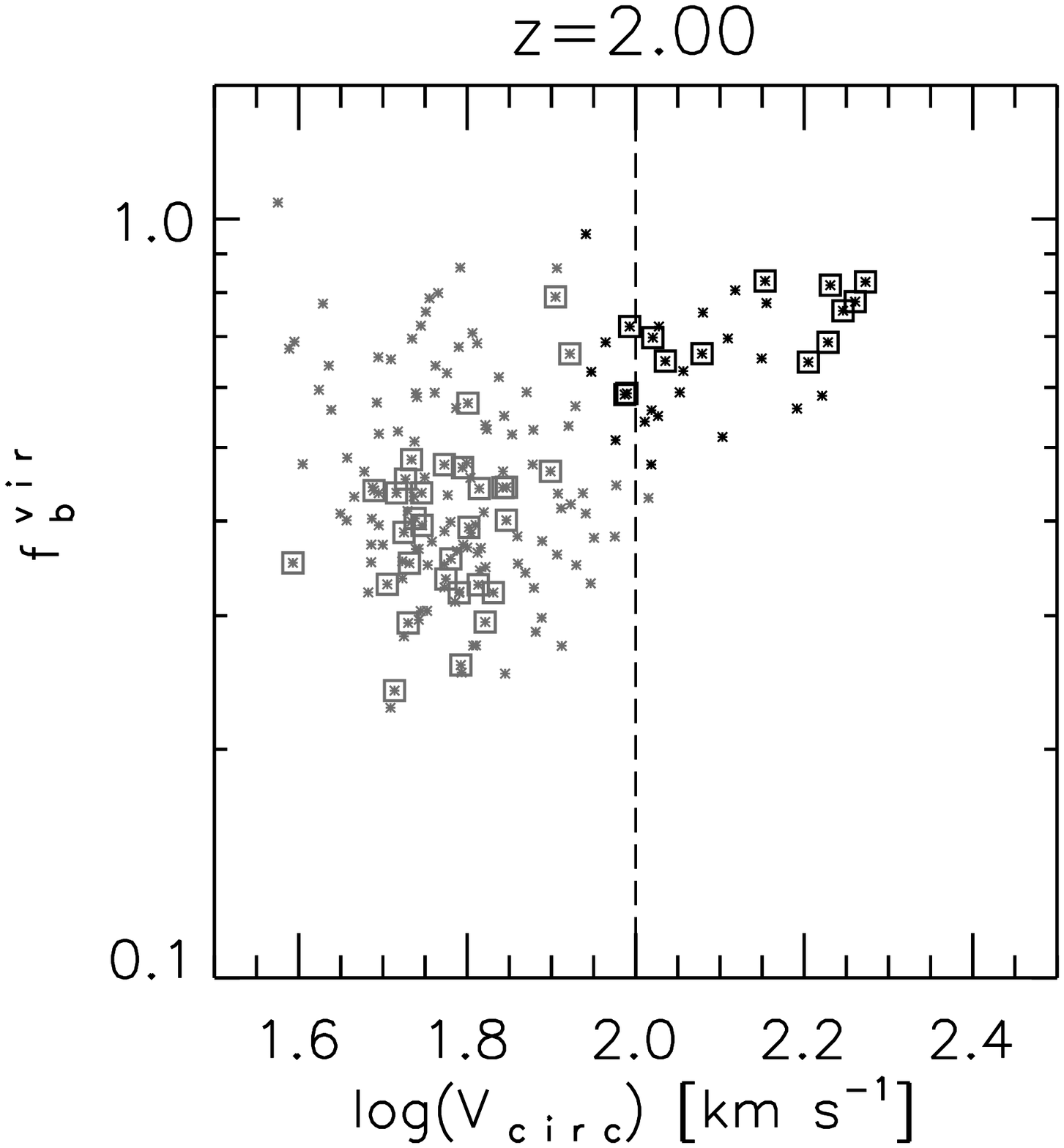}}
\end{center}
\caption{
Left and middle panels:
Fraction of stellar ($f_{*}$) and baryonic ($f_{\rm b}$) 
mass residing in simulated
galaxies relative to the expected baryonic mass within the virial radius
for S230.
Right panel: fraction of the total baryonic mass within the virial radius
relative to the expected one ($f^{\rm vir}_{\rm b}$) for S230.
Results are shown at $z=0$ (upper panels) and $z=2$ (lower panels).
For symbol codes see Fig. \ref{tf}. 
Note that the vertical axes have been plotted
using logarithmic scale.
}
\label{fig:fexpec}
\end{figure*}

To further investigate our findings regarding the origin of the bend of the TFR and the role
played by SN feedback,  we studied 
the fraction $f_{i} = {\Omega}_{\rm m} M_{i}  / {\Omega}_{b} M_{\rm vir}$ 
for each  simulated galaxy at $z=0$, where $M_{\rm vir}$ is the total mass within the virial radius
and $i$ denotes the stellar or the baryonic component within $R_{\rm bar}$.
Hence, the quantity $f_{*}$ ($f_{b}$)
represents the ratio between the stellar (baryonic) mass in a given galaxy and the expected 
baryonic mass within the virial radius inferred from the universal baryonic fraction (${\Omega}_{\rm b}/ {\Omega}_{\rm m}$)
corresponding to the adopted cosmological parameters.
As  can be appreciated from Fig. \ref{fig:fexpec}, $f_{*}$ is an increasing function of the circular velocity which, on its turn, is a measure of the potential well of the systems.
The small systems tend to  have $f_{*} < 0.1$ at $z = 0$, while for more massive galaxies,  $f_{*}$ reaches $\sim 0.3$.
These findings are consistent with the fact that SN outflows are more efficient to heat up and/or expel
the gas content from star forming regions in shallower potential wells, leading to a decrease in their stellar mass content and making them to lay below the sTFR determined by massive galaxies.
At $z =2$, the trends are similar to the local ones with  faster rotators systems exhibiting stellar fractions of around 0.4 and 
most of the slow rotators not exceeding 0.2.
Nonetheless, it is worth noting that, at a given circular velocity, the discrepancies between the simulated stellar components and the 
theoretical baryonic masses
are larger in the local Universe, evidencing the accumulated effects of SN feedback.
We also note that high redshift galaxies show a higher slope for the relation between 
$f_{*}$ and $V_{\rm c}$ indicating the more important action of SN winds on smaller systems at this epoch.

When  the gas mass within $R_{\rm bar}$ is incorporated in the calculations, we obtained the total baryonic fraction $f_{\rm b}$. 
At $z=0$, $f_{\rm b}$ shows a weaker dependence on circular velocities than $f_{*}$. 
More massive galaxies exhibit $f_{\rm b} \sim f_{*}$, indicating that these systems are gas poor. 
On the other hand, small galaxies exhibit baryonic fractions $f_{\rm b}$ which are more than a half larger than $f_{*}$. These correlations explain the trend to 
recover a single slope relation for the local TFR over a larger velocity range when considering  the baryonic content of galaxies instead of their stellar mass. 
However, the fact that $f_{\rm b} < 0.55$ within  $R_{\rm bar}$ for our whole sample indicates that galactic outflows are important over the  whole simulated sample.
The effects on the faster rotators are also indirect since they grow by the accretion
of smaller substructures which are strongly affected by SN feedback as shown before.
 It is worth noting, that the $f_{*}$ and $f_{\rm b}$ predicted by simulations are in qualitative  agreement with observations \citep[e.g.][]{mcg10}, albeit with a weaker dependence on $M_{*}$.
At  $z = 2$, on the other hand, we obtained a stronger correlation between $f_{\rm b}$ and $V_{\rm circ}$. 
Slow-rotating systems at this redshift
show, on average,  similar values of $f_{\rm b}$ to local ones. Conversely,
 fast-rotators in the local Universe
exhibit half of the baryonic fractions derived for systems of similar circular velocities at $z = 2$.  
This behaviour is consistent with the stronger signal for a bend in the bTFR that we found at high redshifts and shows that SN feedback has acted on these systems
expelling part of the baryonic mass outside the very central regions from $z \sim 2$.

In order to estimate if the expelled baryonic mass is within the virial radius
or if SN outflows have been powerful enough to transport it into the intergalactic medium (IGM),
we calculated the
fraction $f^{\rm vir}_{\rm b}$ of the total baryonic mass within the virial radius of simulated galaxies
relative to the one derived from the universal baryonic fraction (${\Omega}_{\rm b}/ {\Omega}_{\rm m}$) in the
adopted cosmology.  The results can
be appreciated in the right panels of Fig. \ref{fig:fexpec}.
By comparing these findings with the ones obtained for $f_{\rm b}$, it
is clear that an important amount of the missing baryons within $R_{\rm bar}$
can be found in the surrounding halo.  However, given that $f^{\rm vir}_{\rm b} < 1$,
there is also a  fraction of the gas which has been  blown away as a consequence
of very  efficient galactic winds. 
These effects are more evident at lower redshifts and for smaller systems. 

We can also see that the correlation with virial mass
is clearly defined at both redshifts albeit with high dispersion.
At $z = 2$, the fraction of absent baryons within $R_{\rm vir}$ 
ranges from 20 per cent to 70 per cent,
with slow rotators having experienced the most important losses, as expected.
In the case of local galaxies, the fraction of lost baryons is within 
the range  30 per cent and 80 per cent.

\begin{figure}\hspace*{0.5cm}\\
\begin{center}
\resizebox{8cm}{!}{\includegraphics{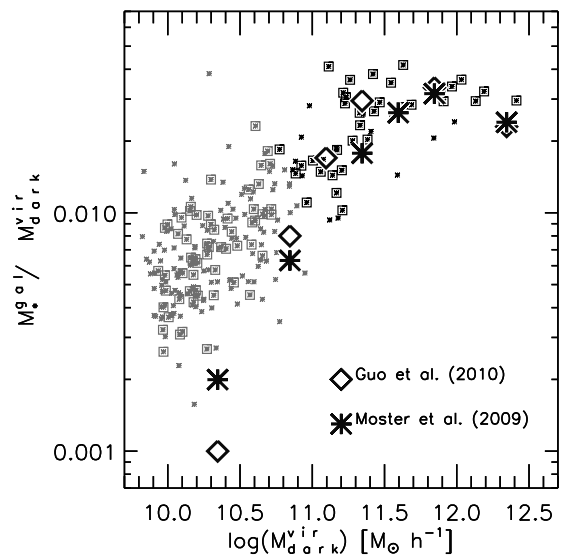}}
\end{center}
\caption{
Fraction of stellar 
mass residing in simulated
galaxies at $z=0$ relative to the dark component  within the virial radius.
For symbol codes see Fig. \ref{tf}.  We show for comparison results from the
semi-analytical works by \citet{guo09} and \citet{moster09}, as indicated in the figure.
Note that the vertical axis has been plotted
using logarithmic scale.
}
\label{fig:fdark}
\end{figure}

Finally, in Fig. \ref{fig:fdark}, we contrast our findings with
the trends obtained from the semi-empirical models of \citet{guo09} and \citet{moster09} 
which are derived from the observed luminosity function.
This figure shows the fraction of stellar mass ($F_{*}$)  
contained within each simulated
galaxy at $z=0$ relative to the dark component  within the virial radius.
We see that there is a very good agreement between our results and previous works.
However, we note that our model still tends to overpredict the stellar mass fractions 
at the low-mass end of the relation.

The dynamical evolution of galaxies in our simulations is driven by the joint action of several physical processes such as the star formation mechanism, gas infall and galactic outflows. 
The star formation process is triggered mainly when the gas gets cold and dense
\footnote {
Recall that in this model the cold phase is defined as the gas component with temperature $T < T_c $ where $ T_c = 8\times 10^4$ K 
and density $\rho > 0.1 \rho_{\rm c}$ where $\rho_{\rm c} $ is $7 \times 10^{-26} {\rm g \ cm^{-3}}$ (Section \ref{sec:simus}).
Otherwise, the gas is classified as hot phase.
}
. 
As a consequence of star formation, SN energy is released heating up the surrounding cold and dense ISM which, when appropriated (see section \ref{sec:simus}), is promoted
to the hot phase. 
This generates  a stronger decrease in the  star formation activity in 
systems with  shallower potential wells because it is easier for the 
cold and dense gas in small systems 
to match the entropy of  the hot and/or diffuse phase residing within their
small dark matter haloes.
 But, the cooling times ($t_{\rm cool}$) of the promoted particles are  
too short compared to their dynamical times ($t_{\rm dyn}$) 
to allow this  hot phase to be stable for a long time
\footnote{In our model, the cooling and dynamical times are calculated following standard definitions 
as $t_{\rm cool}= 3 k T \mu m_{\rm p} / 2 \Lambda \rho$
and  $t_{\rm dyn}= \sqrt{3 \pi / 6 G \rho}$, respectively.
}. Hence, the gas which has just reached the hot phase due to SN heating 
might cool down again in short timescales, returning to the cold phase. Therefore, SN feedback 
leads to a self-regulated cycle of heating and cooling exerting an important influence on the regulation of the 
star formation process in  low-mass galaxies.
 As the systems get larger, the hot phase is established at a higher temperature due to the joint effect of the continuous pumping of SN energy into the hot environment
and the  increase
 of the virial temperature of the dark haloes hosting them.
Hence, as the cooling times get longer compared to the dynamical time scales, the hot gas is able  to remain in this phase.
While the hot environment is able to build up as galaxies get more massive,  
galactic winds would be more difficult to be triggered in these galaxies, since  more SN energy 
would be required by the
cold phase in order to match the entropy of its  nearby hot environment and,  therefore, to be promoted generating outflows. 
Meanwhile, the cold gas
remains available for star formation. 
It is also true that as the systems grows, their gas reservoir decreases, also decreasing their
star formation activity and consequently, the source of SN energy.
Therefore, the action of SN feedback is much less 
important at regulating the transformation of the remaining gas into stars in these large systems.

\begin{figure*}\hspace*{0.5cm}\\
\begin{center}
\resizebox{6cm}{!}{\includegraphics{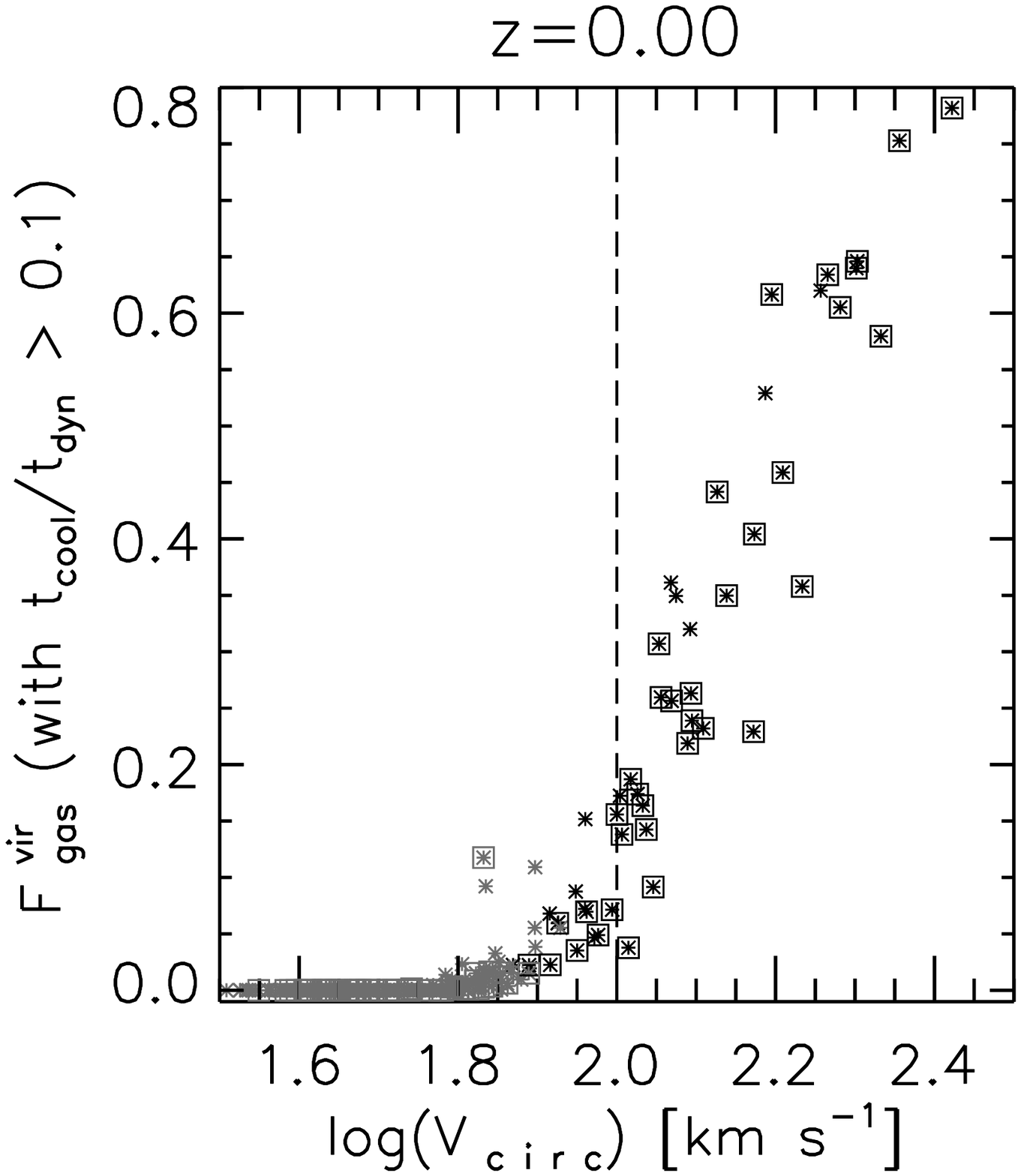}}
\resizebox{6cm}{!}{\includegraphics{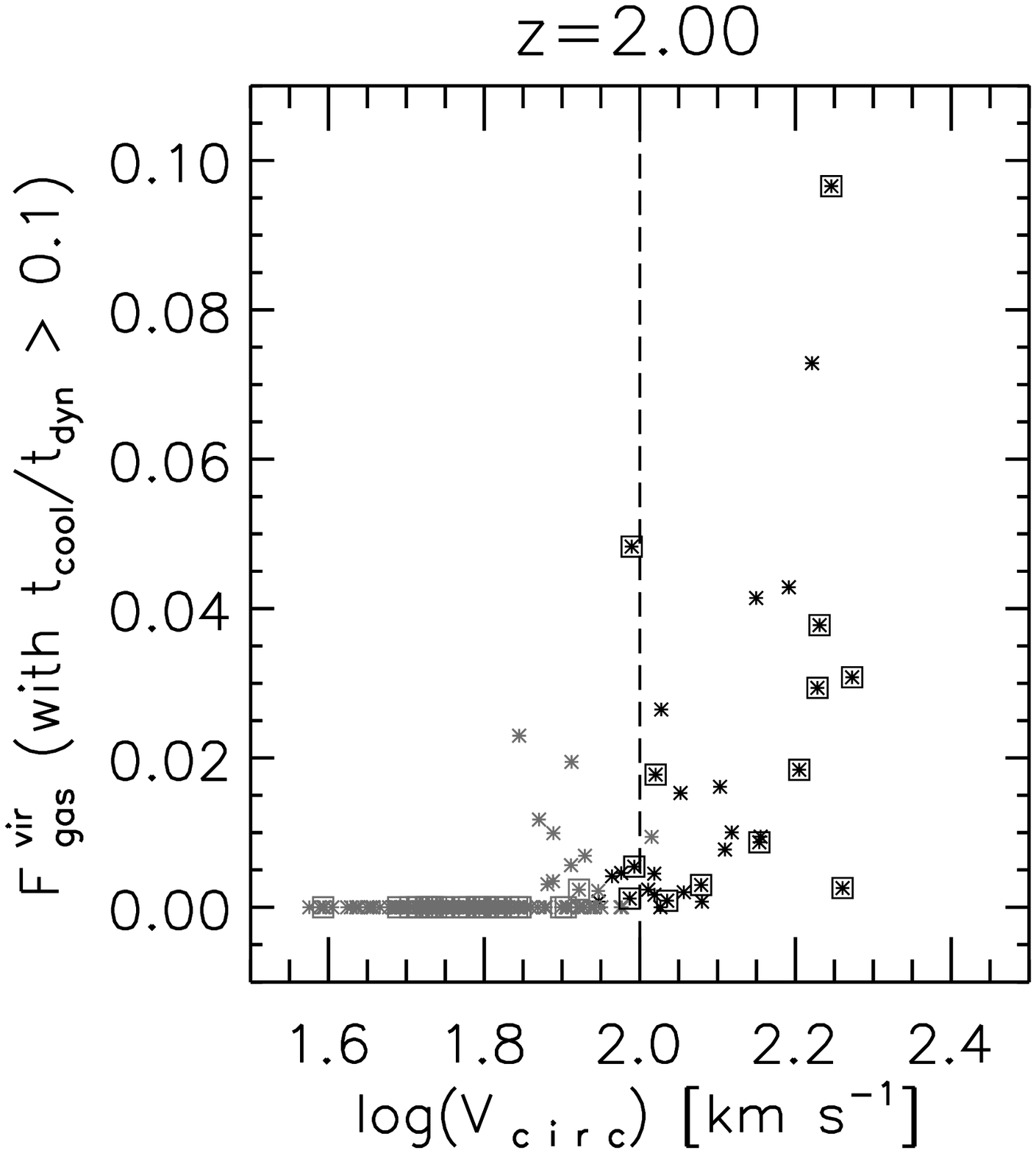}}
\end{center}
\caption{
Fraction of the gas component within $R_{\rm vir}$  which satisfies the condition
$t_{\rm cool}/t_{\rm dyn} > 0.1$ in S230 at $z=0.0$ (left panel) and
$z=2.0$ (right panel).
For symbol codes see Fig. \ref{tf}.  }
\label{fig:tcool}
\end{figure*}

In order to quantitatively probe that these behaviours are taking place in our systems, we estimated the relations between the dynamical  and the cooling  times of the cold gas that has joined the hot phase
by the action of SN feedback (i.e. promoted particles), along the evolution of our simulated volumes.
The fraction of  gas-phase non-available for 
efficient cooling within simulated galaxies ($F^{\rm vir}_{\rm gas}$) is defined as the gas mass within $R_{\rm vir}$
which satisfies the condition $t_{\rm cool}/t_{\rm dyn} > 0.1$ \citep{navarro1993}.
In Fig. \ref{fig:tcool}, we plot $F^{\rm vir}_{\rm gas}$
as a function of $V_{\rm circ}$  in S230 at $z=0$ (left panel) and
$z=2$ (right panel).
As expected, $F^{\rm vir}_{\rm gas}$ is an increasing function of $V_{\rm circ}$.
At $z=0$, systems with $V_{\rm circ} < 100 {\rm \ km \ s^{-1}}$ has an important
fraction of their gas-phase ($> 90 $ per cent) subject to an efficient cooling.
Only at $V_{\rm circ} > 100 {\rm \ km \ s^{-1}}$, $F^{\rm vir}_{\rm gas}$
starts to increase reaching 0.8 at $V_{\rm circ} \sim 250 {\rm \ km \ s^{-1}}$.
The same trends are present at $z =2$, albeit with a larger dispersion at the
high-velocity end.  
 Interestingly,  our model predicts that the transition from efficient to inefficient gas cooling occurs
at around the  characteristic velocity $V_{\rm c} \sim 100 {\rm \ km \ s^{-1}}$ where
the TFR bends. This velocity is also in agreement with
the theoretical expectations from \citet{ds86} who concluded that 
only for systems with $V_{\rm cir}$ larger than  $\sim 100 \ {\rm km \ s}^{-1}$,  
the cooling times of the diffuse and hot gas are at least one order of magnitude longer than the 
dynamical times and, therefore,  the gas could remain hot.

\section{Conclusions}

\label{sec:conclusions}

We have studied the sTFR and bTFR by performing hydrodynamical simulations in a cosmological scenario 
including the action of the combined multiphase and SN feedback model of \citet{scan05, scan06}. 
Our results suggest the existence of a change in the slope of the sTFR and bTFR produced by the 
action of SN feedback.
At $z=0$, the simulated sTFR is very tightly defined  and  in general
 agreement with observations  \citep[e.g.][]{mcg00}.
We detect a systematical change in the slope of the sTFR from  approximately  $\sim 100 \ {\rm km \ s}^{-1}$, at least since $z=3$, so that
lower velocity systems tend to have smaller stellar masses than those predicted by the linear regression of the fast rotators.
The comparison with higher and lower  resolution runs shows that this feature is not just a numerical artefact. 
Regarding the local bTFR, over the velocity range resolved by our simulations, it can be fitted by a single linear model. At higher redshifts, we  detect a weak trend for a change in the slope.
Similar trends are found by using $V_{\rm max}$ as the kinematical estimator.

Our results suggest an evolution of $\approx 0.44$ dex from $z=3$ to $z=0$ for the sTFR  which is global agreement with that reported by \citet{cres09} of $0.41 \pm 0.11$ dex from $z \approx 2.2$ to $z=0$. For the baryonic relation, our model predicts a slightly weaker evolution of  $\approx 0.30$ dex from $z=3$ to $z=0$.
In the wind-free run, both the sTFR and the bTFR show the same level of evolution ($\approx 0.55$ dex) as expected since
baryons are efficiently transformed into stars as soon as they collapse within the potential well of a galaxy. 

The characteristic velocity of $\sim 100 \ {\rm km \ s}^{-1}$ appears {\it naturally } as a consequence of
the physically motivated multiphase and SN feedback treatment adopted in these simulations. 
 This model is successful at establishing a self-regulated
star formation process without using any global properties of the systems but resorting to the local
estimations of the thermodynamical properties of the gas  in the distinct components of the ISM.
We have checked that this result is robust against variations of the SN input parameters.
 Nevertheless, we acknowledge the existence of several issues that need to be solved  in the future such as the level of change in the slope which seems to be weaker in our simulations 
compared to observations. In this regard, our findings suggest that the study of the sTFR and bTFR at different cosmic times could help to constrain SN feedback models.

\begin{acknowledgements}

We thank the anonymous referee for his/her useful comments that largely
helped to improve this paper.
The authors are grateful to Cecilia Scannapieco for making available the S320 simulation and to Tom\'as Tecce for useful discussions. We acknowledge support from the  PICT 32342 (2005) and
PICT 245-Max Planck (2006) of ANCyT (Argentina).
Simulations were run in Fenix and HOPE clusters at IAFE and Cecar cluster at  University 
of Buenos Aires.

\end{acknowledgements}


\begin{thebibliography}{64}
\bibliographystyle{aa}

\bibitem[Avila-Reese, Firmani, \& Hern{\'a}ndez (1998)]{vd98} {Avila-Reese, V., Firmani, C., \&  Hern{\'a}ndez X.} 1998, ApJ, 505, 37

\bibitem[Avila-Reese et al. (2008)]{avila09} Avila-Reese, V., Zavala, J., Firmani, C., \& Hern{\'a}ndez-Toledo, H.~M. 2008, AJ, 136, 1340

\bibitem[Bell \& de Jong (2001)]{bdj01} Bell, E.~F., \& de Jong, R.~S. 2001, ApJ, 550, 212

\bibitem[Conselice et al. (2005)]{cons05} Conselice, C.~J., Bundy, K., Ellis, R.~S.,
Brichmann, J., Vogt, N.~P., \& Phillips, A.~C. 2005, ApJ, 628, 160

\bibitem[Cresci et al. (2009)]{cres09} Cresci, G., et al. 2009, ApJ, 697, 115

\bibitem[Croft et al. (2009)]{croft09} Croft, R.~A.~C., Di Matteo, T., Springel, V., \& Hernquist, L. 2009, MNRAS, 400,
43

\bibitem[De Rijcke et al. (2007)]{rijcke07} De Rijcke, S., Zeilinger, W.~W., Hau, G.~K.~T., Prugniel, P. \& Dejonghe, H.
2007, ApJ, 659, 1172

\bibitem[Dekel \& Silk (1986)]{ds86} Dekel, A. \& Silk, J. 1986, ApJ, 303, 39

\bibitem[Dutton \& van den Bosch (2009)]{vdbosh09} Dutton, A.A. \& van den Bosch, F.C. 2009, MNRAS, 396, 141

\bibitem[Flores et al. (2006)]{flo06} Flores, H., Hammer, F., Puech, M., Amram, P., \& Balkowski, C. 2006, A\&A, 455, 107

\bibitem[Governato et al. (2007)]{gov07} Governato, F., Willman, B., Mayer, L., Brooks,
A., Stinson, G., Valenzuela, O., Wadsley, J., \& Quinn, T. 2007, MNRAS, 374, 1479

\bibitem[Guo et al. (2010)]{guo09} Guo, Q., White, S., Li, C. \& Boylan-Kolchin, M. 2010,
MNRAS, 404, 1111

\bibitem[Gurovich et al. (2004)]{guro04} Gurovich, S., McGaugh, S.~S., Freeman, K.~C.,
Jerjen, H., Staveley-Smith, L., \& De Blok, W.~J.~G. 2004, PASA, 21, 412

\bibitem[Gurovich et al. (2010)]{guro10} Gurovich, S., Freeman, K.~C.,
Jerjen, H., Staveley-Smith, L., \& Puerani, I. 2010, AJ, in prenss (arXiv:1004.4365)

\bibitem[Kang et al. (2005)]{kang05} Kang, X., Jing, Y.P., Mo, H. J., Borner, G. 2005, ApJ,
631, 21

\bibitem[Larson (1974)]{larson04} Larson, R. B. 1974, MNRAS, 169, 229

\bibitem[Mac Low \& Ferrara (1999)]{maclow99} Mac Low, M.M., \& Ferrara, A. 1999, ApJ, 513, 142

\bibitem[McGaugh et al. (2000)]{mcg00} McGaugh, S.~S., Schombert, J.~M., Bothun,
G.~D., \& de Blok, W.~J.~G. 2000, ApJ, 533, L99

\bibitem[McGaugh (2005)]{mcga05} McGaugh, S.~S. 2005, ApJ, 632, 859

\bibitem[McGaugh et al. (2010)]{mcg10} McGaugh, S.~S., Schombert, J.~M., de Blok,
W.~J.~G., \& Zagursky, M.~J. 2010, ApJ, 708L, 14

\bibitem[Mo, Mao, \& White (1998)]{mo98} Mo, H.~J., Mao, S., \& White, S.~D.~M. 1998, MNRAS, 295, 319

\bibitem[Mosconi et al. (2001)]{mos01} Mosconi, M. B., Tissera, P. B., Lambas, D. G., \& Cora, S. A. 2001, MNRAS, 325, 34

\bibitem[Moster et al. (2009)]{moster09} Moster, B.P., Somerville, R.S., Maulbetsch, C., van den Bosch, F.C., Maccio', A. V., Naab, T., \& Oser, L. 2009, ApJ, 710, 903

\bibitem[Nagashima et al. (2005)]{nag05} Nagashima, M., Yahagi, H., Enoki, M., Yoshii, Y., \& Gouda, N. 2005, ApJ, 634, 26

\bibitem[Navarro \& White (1993)]{navarro1993} Navarro, J. F., \& White, S. D. M. 1993, MNRAS, 265, 271

\bibitem[Navarro \& Steinmetz (2000)]{Nav2000} Navarro, J. F., \& Steinmetz, M. 2000, ApJ, 538, 477

\bibitem[Puech et al. (2010)]{puech10} Puech, M., Hammer, F., Flores, H., Delgado-Serrano, R., Rodrigues, M., \& Yang, Y.
2010, A\&A, 510, 68

\bibitem[Raiteri et al. (1996)]{raiteri96} Raiteri, C.~M., Villata, M., \& Navarro, J.~F. 1996, A\&A, 315, 105

\bibitem[Scannapieco et al. (2005)]{scan05} {Scannapieco, C., Tissera, P.B., White, S.D.M. \& Springel, V.} 2005,
MNRAS, 364, 552 

\bibitem[Scannapieco et al. (2006)]{scan06} {Scannapieco, C., Tissera, P.B., White, S.D.M. \& Springel, V.} 2006,
MNRAS, 371, 1125 

\bibitem[Scannapieco et al. (2008)]{scan08} Scannapieco, C., Tissera, P.B., White, S.D.M. \& Springel, V. 2008,
MNRAS, 389, 1137

\bibitem[Springel et al. (2001)]{springel01} Springel, V., White, S.D.M., Tormen, G. \& Kauffmann, G. 2001, MNRAS, 328, 726


\bibitem[Springel \& Hernquist (2003)]{sh03} Springel, V., \& Hernquist, L. 2003, MNRAS, 339, 289

\bibitem[Springel (2005)]{springel2005} Springel, V. 2005, MNRAS, 364, 1105

\bibitem[Steinmetz \& Navarro (1999)]{sn99} Steinmetz, M., \& Navarro, J.~F. 1999, ApJ, 513, 555

\bibitem[Stinson et al. (2007)]{stin07} Stinson, G.~S., Quinn, T., Dalcanton, J.,
Wadsley, J., \& Gogarten, S. 2007, AAS, 38, 766

\bibitem[Tassis et al. (2008)]{tassis08} Tassis, K., Kravtsov, A.V., \& Gnedin, N.Y. 2008, ApJ, 672, 888

\bibitem[Thielemann et al. (1993)]{thiel93}
Thielemann, F.~K., Nomoto, K., \& Hashimoto, M. 1993, in
Origin and Evolution of the Elements, ed. N. Prantzoz et al. (Cambridge University Press), 297

\bibitem[Tissera, Lambas \& Abadi (1997)]{tis97} Tissera, P. B., Lambas, D. G., \& Abadi, M. G. 1997, MNRAS, 286, 384

\bibitem[Tully \& Fisher (1977)]{tf77} Tully, R.~B., \& Fisher, J.~R. 1977, A\&A, 54, 661

\bibitem[Verheijen (2001)]{ver01} Verheijen, M.~A.~W. 2001, ApJ, 563, 694

\bibitem[White \& Frenk (1991)]{wf91} White, S.~D.~M., \& Frenk, C.~S. 1991, ApJ, 379, 52

\bibitem[Woosley \& Weaver (1995)]{WW95}
Woosley, S.~E., \& Weaver, T.~A. 1995, ApJS, 101, 181

\end{thebibliography}
\end{document}